\documentclass[lettersize,journal]{IEEEtran}
\ifCLASSINFOpdf
\else
\fi
\usepackage{booktabs} 
\usepackage[ruled]{algorithm2e} 
\usepackage{framed}
\usepackage{multicol}
\usepackage{listings}
\usepackage[]{algorithm2e}

\usepackage{algpseudocode}
\SetAlFnt{\small}
\SetAlCapFnt{\small}
\SetAlCapNameFnt{\small}
\SetAlCapHSkip{0pt}
\IncMargin{-\parindent}

\usepackage{color}
\usepackage[dvipdfmx]{graphicx}
\graphicspath{{figures/}}
\newtheorem{Definition}{Definition}[section] 

\begin{document}
%
\title{Embedded System Evolution in IoT System Development Based on MAPE-K Loop Mechanism}


\author{
  \IEEEauthorblockN{
    Hiroyuki Nakagawa\IEEEauthorrefmark{1},
    Shinya Tsuchida\IEEEauthorrefmark{1},
    Emiliano Tramontana\IEEEauthorrefmark{2}, 
    Andrea Fornaia\IEEEauthorrefmark{2}, and
    Tatsuhiro Tsuchiya\IEEEauthorrefmark{1}
    }
  \IEEEauthorblockA{\IEEEauthorrefmark{1}
    Osaka University, Yamadaoka 1-5, Suita, Osaka, Japan\\
    \{nakagawa, s-tutida, t-tutiya\}@ist.osaka-u.ac.jp}
  \IEEEauthorblockA{\IEEEauthorrefmark{2}
    University of Catania, Viale A. Doria, 6, Catania, Sicily, Italy\\
    \{tramontana, fornaia\}@dmi.unict.it
  }
}




\IEEEtitleabstractindextext{%
\begin{abstract}
Embedded systems including IoT devices
are designed for specialized functions; 
thus, changes in functions are not considered following their release. 
For this reason, changing functions to satisfy the requirements 
of IoT systems is difficult.
In this study, we focus on updating existing embedded systems without modifying them. 
We investigate the design of new functions and their implementation
with limited resources. 
This paper describes an evolution mechanism for updating the functionalities of existing embedded systems. 
The evolution mechanism uses a control unit that is deployed outside the embedded system.
To guide the steady implementation of the evolution mechanism, 
we define an evolution process that effectively uses the state machine diagram at the design time and runtime to update the embedded systems.
The programming framework implemented in this study supports the evolution process. 
We evaluate the evolution mechanism based on the results from 
two experiments.
The first experiment involved applying the evolution mechanism to a cleaning robot, 
this demonstrated that the evolution mechanism systematically enables the injection of new functions into an embedded system in the real world.
The second experiment, on the probabilistic model checking, 
demonstrated that the mechanism provides almost the same performance 
as the ordinary embedded system with an improved robustness.
\end{abstract}

}

\maketitle

\IEEEdisplaynontitleabstractindextext

%
\IEEEpeerreviewmaketitle

\section{Introduction}

IoT systems are developed by assembling various components, 
including sensors, devices, IoT clouds (software components), 
edge components, and user interfaces. 
Existing {\it embedded systems} are expected to be used as components
of IoT systems for their efficient development.
Embedded systems are usually designed to provide specific services.
To provide these services efficiently, hard limitations
are imposed on the hardware/software of the embedded systems. 
Such limitations prevent the embedded systems from being updated 
after their release. 
%
This characteristic sometimes yields negative effects to the development
of IoT systems.
When developing IoT systems, connecting sensors with devices is often
required to incorporate a new monitoring function \cite{6424332}. 
However, when we use embedded systems as devices in an IoT system,
their lack of flexibility makes it difficult to connect these devices.


\par
In this study, we address the issue related to the difficulty in 
updating embedded systems.
We regard this issue as a problem in the system/software evolution
\cite{Mens08} of embedded systems.
We mainly focus on updating existing embedded systems 
without modifying existing functions and components. 
We use a {\it control loop} to add new functions to embedded systems 
by changing the system control flow. 
The control loop \cite{Kephart03,Oreizy99,Shaw95} is a promising approach
for creating automated systems. 
In particular, the MAPE-K loop \cite{IBM05}
has  attracted attention owing to its autonomous management of 
information systems \cite{Weyns13}. 
We expect that the MAPE-K loop mechanism to be compatible 
with IoT activities at various levels, 
namely monitoring, controlling, optimization, and autonomy 
\cite{porter2014smart}. 

This paper describes an evolution mechanism that supports for updating 
the functionalities of embedded systems.
The evolution mechanism uses a control unit, 
which is deployed outside the embedded system.
The control unit is constructed based on the MAPE-K loop.
A programming framework developed in this study 
aids in constructing the control unit.
We define an evolution process that 
supports the system evolution from the early stage of the 
system design phase to the implementation phase. 
The evolution process uses the state machine model to detect 
and handle changes between the original and new versions of the system.
The state machine models of the two versions are used 
by a MAPE-K loop component at runtime.

\par
The main contributions of this paper are as follows:
\begin{itemize}
 \item[(1)]  {\it Evolution mechanism for embedded systems}:
The mechanism enables the evolution of embedded systems
without modifying embedded systems. 
This mechanism uses an event converter 
to control an embedded system 
to behave as a new embedded system. 
 \item[(2)]  {\it Evolution process}: 
This process systematically implements the evolution of embedded systems
using the evolution mechanism.
This process uses the state machine models to design new functions
and implement evolution to the mechanism.
%
 \item[(3)] {\it Programming framework}:
The framework aids in implementing the evolution in embedded systems 
more steadily. 
In particular, the framework helps us construct
the event converter by providing useful APIs.
This framework is developed based on our previous 
framework \cite{TsuchidaCOMPSAC18}. 
\item[(4)]
{\it Experimental evaluation}:
We apply the mechanism, process, and framework to 
a cleaning robot evolution scenario in the real world
to evaluate the effectiveness of the evolution process.
We also evaluate the performance and robustness of the proposed
mechanism using the probabilistic model checking technique.
\end{itemize}

The remainder of this paper is organized as follows:
Section \ref{sect:background} presents the background of this study, 
including 
the state machine diagram and MAPE-K loop. 
Sections \ref{sect:approach} to \ref{sect:approach_implementation} 
describe the design, implementation, and addition of new functions to 
embedded systems using an evolution example of a lighting bulb system. 
Section \ref{sect:experiment} reports the results of 
applying the framework to a cleaning robot evolution scenario
and verifying the performance and robustness of the proposed evolution
mechanism.
%
Section \ref{sect:discussion} discusses of our approach 
based on the results, 
Section \ref{sect:relatedwork} summarizes related work,
and Section \ref{sect:conclustion} 
concludes this paper and outlines future work directions.

\section{Theoretical Background}\label{sect:background}

\begin{figure}
  \centering
  \includegraphics[width=8.0cm]{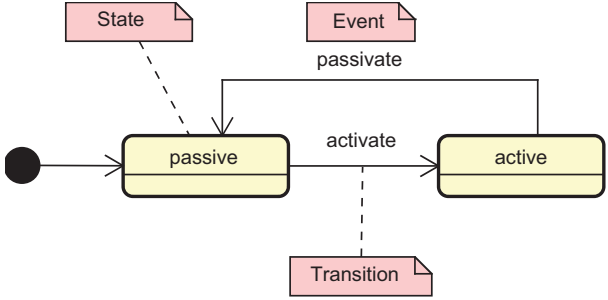}
  \caption{Example of a state machine model. 
The elements labeled ``passive'' and ``active'' correspond to {\it states}, 
the arrows correspond to {\it transitions}, and the labels ``passivate'' and ``activate'' on the transitions are {\it events}.}
  \label{sampleSMD}
\end{figure}

\subsection{State Machine Diagram/Model}

A state machine diagram is a graphical model in the Unified Modeling Language (UML) \cite{UML}. 
The basic state machine diagram is a finite automaton in computer science.
Figure \ref{sampleSMD} presents an example of a state machine diagram,
whose main elements are {\it states} and {\it transitions}.
The state machine diagram forms a graph consisting of states and 
transitions that connect the states.
A state is a situation in the life cycle of an object, 
whereas a transition represents the movement from one state to another. 
Each transition can be labeled as an {\it event} that causes a transition.

The state machine diagram is used extensively in embedded system 
development \cite{Samek08}. 
A main reason for this is that the event-driven architecture provided by
the state machine diagram allows us to describe more flexible patterns
of control than any sequential system \cite{RumbaughBook91}.
The diagram can explicitly define the handling of events that occur 
in each state.
Unlike static UML diagrams including the class diagram, 
the state machine diagram can represent the dynamic behaviors 
of the system.
%
Hereafter, we refer to a model described by the state machine 
diagram as a {\it state machine model}.

\subsection{MAPE-K Loop}

\begin{figure}
  \centering
  \includegraphics[width=7.5cm]{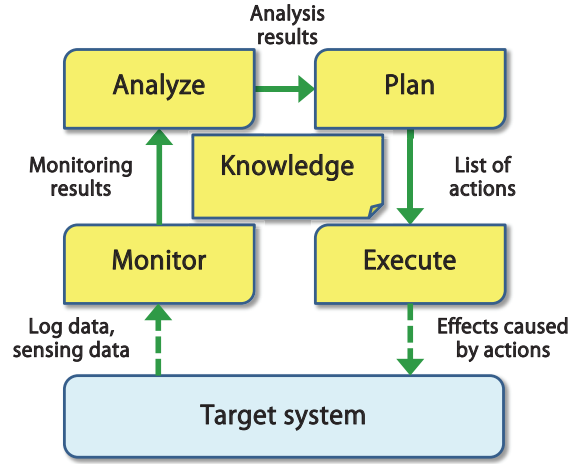}
  \caption{Overview of MAPE-K loop mechanism. The MAPE-K loop manages and controls the target system.}
  \label{mape}
\end{figure}

\begin{figure*}[tpb]
  \begin{center}
  \includegraphics[width=13cm]{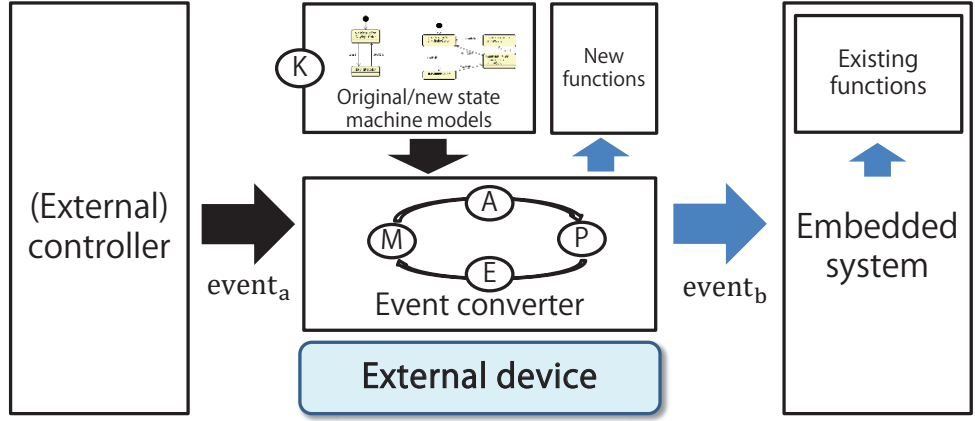}
  \caption{System architecture of our evolution mechanism
that can update the functionalities of embedded systems. 
This architecture uses an event converter deployed in an external device to translate events sent from the controller.}
  \label{approach_1}
  \end{center}
\end{figure*}

The MAPE-K loop \cite{IBM05} was originally developed as a mechanism 
for autonomous software systems, such as self-adaptive systems.
Figure \ref{mape} overviews the MAPE-K loop mechanism \cite{IBM05}.
The mechanism aims to control a software system by continuously executing 
four steps (components), known as {\it monitor, analyze, plan,} and 
{\it execute}, which form a loop. 
A target system is monitored to determine whether problems 
have occurred using logs and sensors at the monitor step.
If a problem is identified at the analyze step, the mechanism attempts 
to determine the cause of the problem. 
Thereafter, at the plan step, actions are planned to solve the problem 
according to the analysis results derived from the analyze step. 
Finally, the actions planned in the previous step 
are performed at the execute step. 
Subsequently, the MAPE-K loop mechanism repeats the activities 
by monitoring the results of the execute step. 
The MAPE-K loop also includes a complementary component known as 
{\it knowledge}, which manages the data to be shared 
among the four components. 
The shared knowledge contains data such as topology information, 
historical logs, and policies.
The aim of this type of loop structure is to identify and handle 
system problems efficiently.

\section{Embedded System Evolution}\label{sect:approach}


This paper describes an embedded system evolution mechanism
that can update the functionalities of embedded systems.
Owing to their hardware constraints and enhanced reliability, 
embedded systems generally exhibit less flexibility for evolution.
To overcome this problem, we adopted an external approach using
additional hardware for the evolution.
Figure \ref{approach_1} depicts the system architecture of 
the evolution mechanism.
Using an external approach that uses additional hardware
(the {\it external device} in Figure \ref{approach_1}), 
additional memory or storage space for implementing new functions 
to be added can be obtained. 
In this mechanism, the event converter plays the central role in 
controlling the new embedded system.
We focus on the event-driven structure of the embedded system.
The event converter changes incoming events sent from
the controller appropriately to control the behavior of the embedded system 
and satisfy the new requirements.

We use the MAPE-K loop structure to construct an event converter.
The MAPE-K loop monitors incoming events and handles the events 
properly according to 
the original and newly updated state machine models. 
The MAPE-K loop enables the handling of events automatically 
and asynchronously.
A programming framework can be used to
construct the event
converter based on the MAPE-K loop structure. 
%
%
To guide the implementation of the evolution mechanism, 
we also define the evolutionary process, 
which encompasses the design and implementation phases of
the development.
The state machine model is the key model in the evolution process.
The model is compatible with the design and 
implementation method of updating existing embedded systems.


\par


\begin{figure}[t]
  \begin{center}
  \includegraphics[width=9cm]{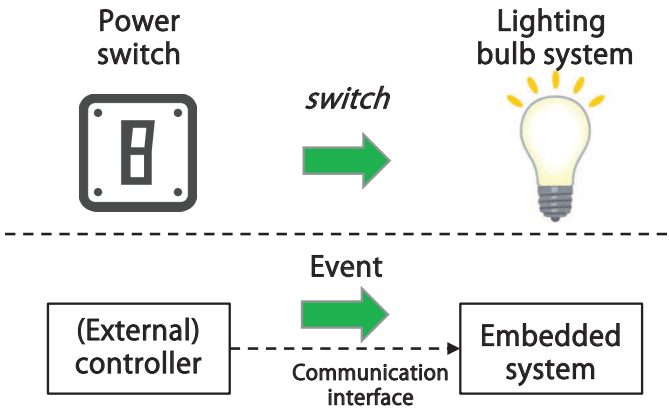}
  \caption{Application domain of light bulb system.}
  \label{light_overview}
  \end{center}
\end{figure}

We consider the evolution scenario of a light bulb system to explain
the evolution process.
Figure \ref{light_overview} depicts the application domain of a 
light bulb system.
The embedded system, i.e., the light bulb system, 
receives commands corresponding to events from an (external) controller, 
which is a power switch in this scenario. 
The system responds to the commands (events) from the controller. 
The communication interface sends/receives events between 
the controller and embedded system using physical media, 
such as ethernet, serial cables, radio communication, and infrared rays.
The controller, which is a power switch, switches light off and on.
Figure \ref{light_original} presents the state machine model of the 
light bulb system.
The light bulb system has two modes: {\it off} and {\it on}. 
When the system receives a switch signal as an event,
the state moves from {\it off} to {\it on}, and vice versa.

We envision the following evolution scenario.
\begin{framed}
{\bf Evolution (add a new color mode):} 
The lighting system should turn on the light in 
the daylight color, which is implemented in the initial version, 
and in the incandescent lamp color, which is a new color tone.
The colors can be changed by pushing the same switch.
The daylight color light can be turned off 
without providing the incandescent lamp color by waiting
two seconds after pushing the switch.
\end{framed}

\noindent
Although this evolution scenario is simple, it requires the fundamental 
problem encountered when updating the functionalities of 
existing embedded systems. 
Thus, new functions must be designed while combining the original 
and new functions.

\section{Design Phase}\label{sect:approach_design}
This section describes the design of the evolution of 
an embedded system in our mechanism.
In our evolution process, new functions for embedded systems
are designed using state machine models.
This section also explains how to handle the changes
using the MAPE-K loop mechanism. 

\subsection{Design Using State Machine Model}\label{subsect:designStep}
\begin{figure}[t]
  \begin{center}
  \includegraphics[width=3.5cm]{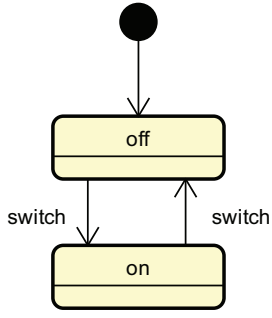}
  \caption{State machine model for light bulb system.}
  \label{light_original}
  \end{center}
\end{figure}

As this study focuses on the evolution of embedded systems,
we assume that embedded systems cannot be designed from scratch.
Furthermore, we assume that a state machine model is available 
for the current version of the system.
If the state machine model does not exist,
it should be constructed for the original (current) version of the system.
Figure \ref{light_original} presents the state machine model for 
the original version of the light bulb system in our example. 
We revise this model to incorporate new functions.

As modifying the equipped functions of embedded systems is difficult, 
the existing states that are implemented,
such as the functions associated with 
the {\it off} and {\it on} states in the light bulb system,
should not be changed.
These states are generally related to hardware resources. 
If the existing states are modified, the part to be modified 
must be identified at the software layer and 
hardware layers. 
This is one of the most difficult tasks in the evolution of 
embedded systems. 
Therefore, we avoid the modification of existing functions
and associated states.

During the design phase, we focus on the event-driven structure.
Most embedded systems are event driven \cite{Samek08}, 
meaning that they continuously wait for internal or external events, 
which may include time triggers and user actions such as button pushes.
The event converter of our evolution mechanism 
extracts incoming events and creates new events. 
To construct the event converter, 
new events and states should be identified
and added to the state machine model. 
The new events and states must be added to the original state machine 
model without removing existing states.
New functions should be described by adding states and transitions
for reaching the states. 
Such modification enables new functions to be added to the embedded system 
without changing the original version.

In a state machine model, a system ({\it Sys}) is represented using 
a finite number of events ($E$), states ($S$), and transitions ($T$). 
The next state is defined using the current state and an upcoming event.
This transformation for defining the next state is known as 
a {\it transition}, which is a member of $T$. 
The system can be represented as a three-tuple:
\begin{eqnarray}
{\it Sys} &=& ( E, S, T ),  \nonumber
\end{eqnarray}
where {\it E, S,} and {\it T} are expressed as follows:
\begin{eqnarray}
 E &=& \{ {\it event_{1}, event_{2}, ... , event_{m}}\}  \nonumber \\
 S &=& \{ {\it state_{1}, state_{2}, ... , state_{n}}\}  \nonumber \\
 T &=&  S \times E \rightarrow S   \nonumber 
\end{eqnarray}

\noindent
Using this notation, the state machine model for the original light 
bulb system ${\it Sys_{o}}$ can be described as follows:

\begin{eqnarray}
\label{eq:ourProcess2}
{\it Sys_{o}} &=& ( {\it E_{o}, S_{o}, T_{o}} ) \nonumber \\
 E_{o} &=& \{ {\it switch}\}  \nonumber \\
 S_{o} &=& \{ {\it off, on} \}  \nonumber \\
 T_{o} &=& \{ ( {\it off, switch} ) \rightarrow {\it on}, ( {\it on, switch} ) \rightarrow {\it off}  \}  \nonumber 
\end{eqnarray}

We evolve the light bulb system to react to the new requirements 
described in Section \ref{sect:approach}.
The original system has two states: {\it off} and {\it on}.
To handle the new requirements, a new function for providing 
a new color tone must be added, 
which should be invoked by pushing the switch.

\begin{figure}[htpb]
  \centering
  \includegraphics[width=9cm]{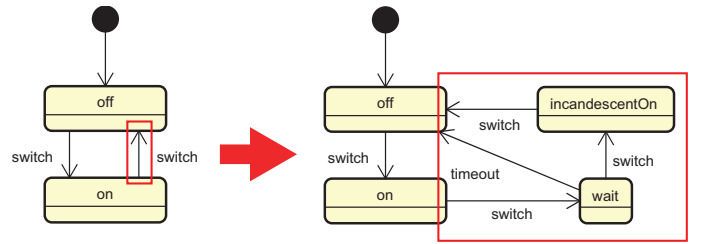}
  \caption{Changes in state machine model of lighting bulb system.}
  \label{light_original_to_new}
\end{figure}

As new functions must be added without modifying the existing states,
events, transitions, and new states are added to the original 
state machine model to construct the new state machine model.
A solution for our evolution scenario is to add two new states, 
{\it wait} and {\it incandescentOn}, to the model.
The former state provides a conditional branch for 
whether the system should change the light to the incandescent color 
or should turn the light off, 
whereas the latter represents the state to provide the incandescent 
color light.
Figure \ref{light_original_to_new} depicts the changes 
in the state machine model that is used to handle the evolution.
The right model in Figure \ref{light_original_to_new} represents 
the new state machine model. 
The {\it timeout} event embedded in the model allows users to 
select whether the light is changed to
an incandescent color or is turned off.
In our example, two seconds after the state is changed to {\it wait},
the internal timer executes this event. 

We define two conditions that are imposed on the new state machine model.
Both conditions guarantee that the new system does not destroy the original
functions related to the individual states and interface of the embedded system.

\noindent
\begin{Definition}\label{defSufficiency}
{\it Necessary conditions of new state machine model: }
The new state machine model should have a set of events ($E_{n}$) and
states ($S_{n}$) that satisfy both of the following conditions.

\begin{itemize}
\item {\it Condition 1:}
$ E_{o} \cap E_{n} = E_{o} $

\item {\it Condition 2:}
$ S_{o} \cap S_{n} = S_{o} $

\end{itemize}
\end{Definition}

\noindent
These conditions guarantee that the states and events of the original 
system are not destroyed. 
New states and events should be added if necessary.
However, transitions can be added as well as modified 
from the original state machine model.
In our example, the new light bulb system ${\it Sys_{n}}$
can be defined as complying with the following conditions:
\begin{eqnarray}
\label{eq:ourProcess3}
{\it Sys_{n}} &=& ( E_{n}, S_{n}, T_{n} ) \nonumber \\
 E_{n} &=& \{ {\it switch, timeout}\}  \nonumber \\
 S_{n} &=& \{ {\it off, on, wait, incandescentOn} \}\nonumber\\
 T_{n} &=& \{ ( {\it off, switch} )  \rightarrow {\it on}, 
( {\it on, switch}) \rightarrow {\it wait}, \nonumber \\
 && ( {\it incandescentOn, switch}) \rightarrow {\it off},  \nonumber \\
 && ( {\it wait, timeout}) \rightarrow {\it off}, \nonumber \\
 && ({\it wait, switch}) \rightarrow {\it incandescentOn} \} \nonumber \\
\end{eqnarray}

\subsection{Design of Event Converter}
\label{subsect:detectAndHandleChanges}


As illustrated in Figure \ref{approach_1}, the event converter 
plays a central role in the evolution mechanism.
We focus on the event-driven structure of embedded systems. 
As indicated in Figure \ref{light_overview},
events are usually sent from the controller to the embedded system.
In our evolution mechanism, 
the events sent from the controller are intercepted 
by the event converter.
The event converter, which is constructed based on the 
MAPE-K loop mechanism, generates a new event that is determined 
from the state machine models and event that the converter 
received.
The converter may invoke a new function implemented on 
the external device instead of generating a new event.

When an existing function of the embedded system should be performed,
the converter sends events to the embedded system to invoke
the existing function.
However, the converter should not send events 
to the embedded system when a new function implemented on 
the external device should be performed. 
Therefore, the event converter must determine 
whether the new system provides an existing or new function.
The converter 
generates events to be sent to the embedded system if necessary.
The event converter uses two state machine models, the 
original and new state machine models, to generate events.
As this converter should behave independently from the controller 
and embedded system, we construct a converter based on the MAPE-K loop. 
In the following, we explain how events are generated 
using the MAPE-K loop mechanism.

The components of the MAPE-K loop have individual tasks, as follows:
\begin{itemize}
\item {\it Knowledge}: includes the original and new state machine models
  and manages their current states.
\item {\it Monitor}: receives events sent from the controller.
\item {\it Analyze}: verifies whether the coming event is acceptable
at the current state.
\item {\it Plan}: generates a plan, which consists of a list of events 
for invoking existing functions and commands to invoke new functions
if necessary.
\item {\it Execute}: sends events to the embedded system or invokes new functions.
\end{itemize}


\begin{algorithm}
 \label{alg:eventConverter}
 \KwResult{The next action is executed}
 \KwData{{\it oModel, nModel}: State machine models for original and new systems //Knowledge}
 \KwData{Current states, ${\it oState}$, $nState$: current states of both models// Knowledge}
\nl \While{true}{
\nl // Monitor events\;
\nl \If{$event_a$ is observed}{
\nl  //Analyze whether the event is acceptable at the current state\;
\nl    \eIf{nModel.existTransition($nState$, $event_a$)}{
\nl  //Analyze the type of event\;
\nl         \eIf{oModel.existState(nModel.getNextState($nState$, $event_a$))}{
\nl          {\it   $nextAction \gets $ oModel.getEvents($oState$, 
nModel.getNextState($nState$, $event_a$));} //Plan\;
\nl            {\it $oState \gets $ oModel.getNextState($oState$, $event_a$)\; }
\nl            {\it $nState \gets $ nModel.getNextState($nState$, $event_a$)\;}
         }{
\nl            {\it $nextAction \gets $ nModel.getTransition($nState$, $event_a$);} //Plan\;
\nl            {\it $nState \gets $ nModel.getNextState($nState$, $event_a$);}
         }
    }{
\nl            $nextAction \gets none$; 
    }{
\nl {\it execute(nextAction);} //Execute
    }
}{
}
}
\caption{Behavior of event converter.}
\end{algorithm}

The event converter determines the new actions according 
to Algorithm \ref{alg:eventConverter}.
If the monitor component observes the event occurrence (line 3),
the analyze component probes whether the event is acceptable at the current state (line 5). 
The analyze component determines whether the system should use 
an existing or new function (line 7). 
The plan component plans the next action based on the analysis result
(lines 8 and 11). 
Finally, the execute component performs the action determined by 
the plan component (line 14) and updates each state to the next state
 (lines 9, 10, and 12).

For the light bulb system, the initial state is the {\it off} state 
in both the original and new models.
If a $switch$ event occurs, the event converter generates 
a $switch$ event.
The next state, i.e., the {\it on} state, which is the return value 
of the {\it nModel.getNextState(off, switch)} method, 
exists in both the original and  new state machine models 
(Figure \ref{light_original_to_new}).
As the state exists in both models,
the converter uses an existing function of the embedded system.
In this case, the event converter generates the same event 
as the received event without changing it 
(${\it event_b = event_a}$ in Figure \ref{approach_1}).
We consider the situation in which the $switch$ event occurs 
at the {\it incandescentOn} state in the new model.
The next state, i.e., the return value of the
{\it nModel.getNextState(incandescentOn, switch)} method,
is the {\it off} state.
The {\it off} state exists in both the original and new models.
To use a function provided by the existing embedded system,
the converter must send events to the system to change 
the system state; that is, the original model state.
The current state in the original model is the {\it on} state.
Therefore, a $switch$ event is sent to the embedded system,
and the state is changed from {\it on} to {\it off};
that is, the converter sends $event_a$ (the {\it switch} event) 
as $event_b$ to the embedded system without any changes.

Next, we consider the situation in which the $switch$ event occurs 
at the {\it on} state in the new lighting bulb system.
The next state is {\it wait}, which is the return value of 
the {\it nModel.getNextState(On, switch)} method. 
This state does not exist in the original model, 
but it exists in the new model (Figure \ref{light_original_to_new}).
In this case, new functions are used on the external device.
Therefore, the event converter invokes new functions that are 
associated with the {\it wait} state without generating any events, 
and updates the current state of the new model. 

When the {\it timeout} event arrives at the {\it wait} state 
in the new light bulb system,
the next state, the return value of  
{\it nModel.getNextState(wait, timeout)}, is the {\it off} state. 
As the state exists in both models, 
the existing functions of the embedded system associated with the
{\it off} state are used.
The converter sends the {\it switch} event to the system
to use the functions provided by the existing embedded system. 
%

\section{Implementation Phase}\label{sect:approach_implementation}
This section describes the programming framework for implementing
the event converter based on the MAPE-K loop mechanism.
We explain the implementation of a new function corresponding
to a new state in the new state machine model
and the use of the functions from our framework.

\subsection{Event Converter Implementation}\label{subsect:framework}

\subsubsection{State Machine Models}
The event converter uses two state machine models as behavioral models
to extract and handle changes between the original and updated systems.
Our framework assumes that state machine models are constructed 
using Astah \cite{Astah}, which is a UML modeling tool.
Astah can export models into the XML metadata interchange (XMI) format, which aids in exchanging UML models across different UML tools.
By parsing state machine models in XMI format,
the event converter can identify the events to be detected so that
it can determine a transition for the next state.

\subsubsection{External Device}
Our evolution mechanism uses an external device 
in which the event converter is deployed.
The following mandatory requirements apply to the external device:

\begin{enumerate}
\item The device should have a communication mechanism with the embedded 
system such that the event converter can send events to the embedded system.
 \item The device should be as small as possible to satisfy 
the physical constraints of the embedded system. 
\end{enumerate}

To satisfy these requirements, 
we deploy an event converter on a Raspberry Pi \cite{raspi}. 
Raspberry Pi is a small device; however, 
because it can use an operating system such as {\it Ubuntu}, 
various programs can run on it. 
It also has several interfaces such as GPIO pins and USB ports, 
to communicate with other devices. For these reasons, 
Raspberry Pi is a suitable device for deploying an event converter. 

\begin{figure}[t]
  \centering
  \includegraphics[width=8.6cm]{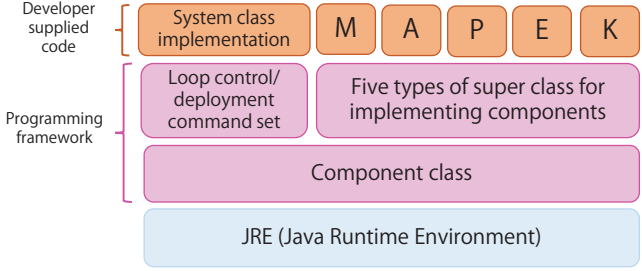}
  \caption{Programming model. The labels ``M'', ``A'', ``P'', ``E'', 
 and ``K''
    represent the monitor, analyze, plan, execute, and knowledge components,
    respectively.}
  \label{Figure:programming}
\end{figure}

\begin{figure}
  \centering
  \includegraphics[width=7.75cm]{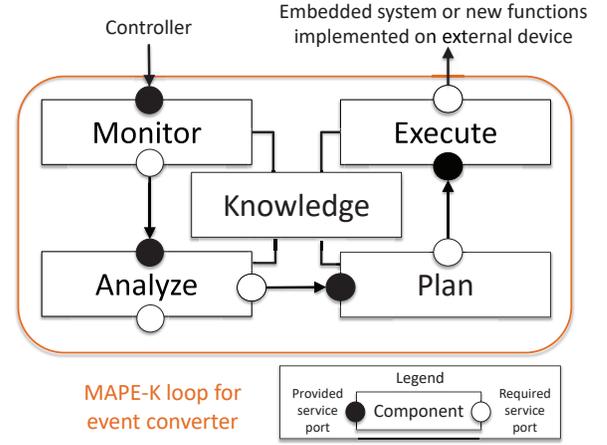}
  \caption{Architectural configuration of the event converter.
Arrows represent the data flow.}
  \label{Figure:component}
\end{figure}

\subsubsection{Programming Framework for Event Converter}
Our programming framework aids in implementing the event converter
based on the MAPE-K loop mechanism.
The framework provides APIs for developing MAKE-K loop mechanisms, 
as indicated in Algorithm \ref{alg:eventConverter}.
A developer mainly implements two parts using this framework:
the code for monitoring events sent from the controller, 
and the code for sending events to the embedded system 
or for calling new functions.
Thus, the two parts correspond to the interfaces of the event converter.

We use Java to execute multiple components in parallel, 
as it is a language that provides multi-thread programming.
We previously implemented a lightweight programming framework 
for operations on real-world hardware \cite{Tsuda15} \cite{Nakagawa12}.
This framework assumes that a system is developed based on 
a component-based structure \cite{4221625}, 
in which components can be added, removed, and replaced dynamically.


\begin{figure*}[t]
\begin{center}
\includegraphics[width=14cm]{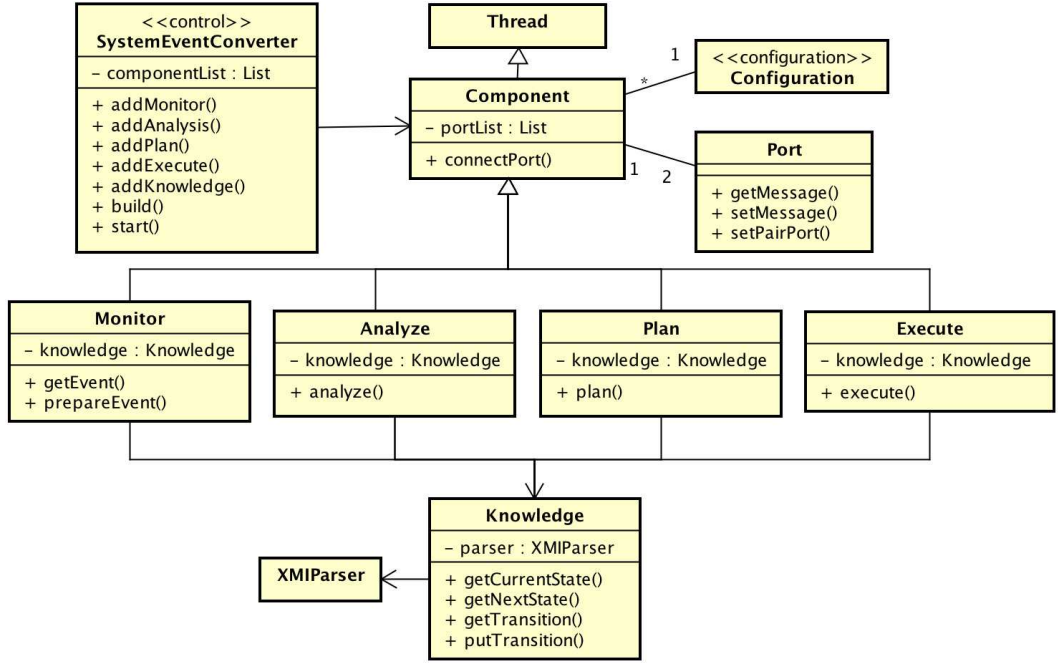}
\caption{Class diagram of programming framework.}
\label{Figure:class}
\end{center}
\end{figure*}

Figure \ref{Figure:programming} presents the programming model of 
the proposed framework. 
This framework provides the concurrent execution of classes, 
which are implemented as extensions of the Java {\it Thread} class.
Figure \ref{Figure:component} depicts the architectural configuration 
of the event converter, 
which is implemented based on the MAPE-K loop mechanism.
The knowledge component is connected to the other components 
to share the data, including the state machine models.
Figure \ref{Figure:class} displays a class diagram 
of the programming framework.
We introduce two main groups of classes that provide useful APIs.
The first group consists of super-classes for implementing 
five types of MAPE-K loop components.
We can focus on implementing individual concerns 
by inheriting these super-classes. 
The second group contains the 
{\it SystemEventConverter} class, which controls
the MAPE-K components implemented 
by inheriting one of the super-classes belonging to the first group.
Each {\it monitor, analyze, plan,} and {\it execute} component generally 
starts its activity after the previous component finished its activity.
The threads of their components are controlled 
by the {\it SystemEventConverter} class, 
in the order of {\it monitor, analyze, plan,} and {\it execute}.
Once the {\it execute} thread completes, 
the MAPE-K loop moves repeatedly from the {\it monitor} thread.
The {\it SystemEventConverter} class handles thread processing 
and data flow, as illustrated in Figure \ref{Figure:component}. 
By inheriting and using these classes, 
a developer can implement the event converter 
without considering the thread synchronization 
and data flow of the MAKE-K loop process.

The event converter is deployed on the external device. 
The framework provides a command set for deploying the event converter with
the MAPE-K loop components.
Table \ref{tbl:command} lists commands that are provided.
Among these commands, the control commands, 
such as \texttt{start} and \texttt{stop}, 
change the execution state of the MAPE-K loop components,
whereas the \texttt{status} command indicates the current status 
of the MAPE-K loop.

\begin{table}[ht]
\centering
\caption{Deployment/Control Command Set}
\small 
 \begin{tabular}{c|l} 
  \hline
  Command& Description \\ \hline \hline 
  \texttt{start} & Start the MAPE-K loop process. \\ \hline
  \texttt{stop} & Stop the MAPE-K loop process. \\ \hline
  \texttt{status} & Show the current status of the MAPE-K loop. \\ \hline
  \texttt{exit} & End the MAPE-K loop process. \\ \hline
 \end{tabular}
 \label{tbl:command}
\end{table}

\subsubsection{Development Steps} \label{sect:imp_eventconverter}
The event converter is implemented and deployed 
according to the following steps:

{\bf (1) Behavior coding:}
The components of the event converter should be implemented
as MAPE-K components.
These components can be implemented by extending the super-classes 
and overriding the methods in the super-classes provided by the framework. 
For example, the class for event monitoring can be implemented by 
inheriting the {\it Monitor} super-class 
and overriding its {\it getEvent} method.
The {\it getEvent} method should end by returning an event 
that is handled by the method. 
The result of the monitor component process is passed to 
the analyze component as an argument for the {\it analyze} method. 
The processes of the analyze and plan components, 
which are indicated in Algorithm \ref{alg:eventConverter}, 
have already been implemented in parent classes; 
therefore, we do not need to change these classes.
New functions corresponding to new states in the new state machine model 
should be implemented in this step.

\begin{figure}[htpb]
  \centering
  \includegraphics[width=9.85cm]{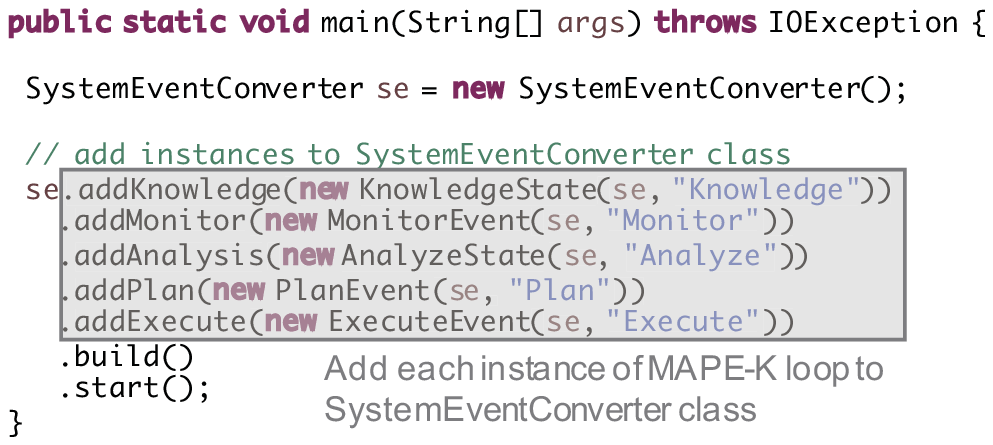}
  \caption{SystemEventConverter class, which provides the loop control 
of the MAPE-K loop.}
  \label{code:MAPE-K}
\end{figure}

{\bf (2) Component configuration:}
The instances of the implemented components 
are registered in the list defined in {\it SystemEventConverter} class 
using {\it add*} methods (Figure \ref{code:MAPE-K}).
After registering the instances, 
the {\it build} and {\it start} methods connect components 
and start the MAPE-K loop, respectively. 
In Figure \ref{code:MAPE-K}, the {\it MonitorEvent} class is a subclass of 
the {\it Monitor} parent class. 
An object of the {\it MonitorEvent} class is registered to 
the {\it SystemEventConverter} object using the {\it addMonitor} method.

{\bf (3) Deployment of event converter:} 
New functions corresponding to new states in the new state machine model 
should also be registered. 
These new functions are called by a method that is included in 
the Knowledge class (Section \ref{sect:imp_newfunctions}). 
After creating an executable file, which is a jar file,
developers deploy the event converter and control the MAPE-K loop 
by executing the commands listed in Table \ref{tbl:command}.

\subsection{Implementation of New Functions}\label{sect:imp_newfunctions}
If a function corresponding to a new state is simple and not large,
the function should be implemented as a single class. 
Otherwise, we first break the corresponding state into sub-states;
that is, the state is replaced with a composite state. 
In this case, we implement each sub-state as a {\it concrete state} 
in the state pattern of the design patterns \cite{Gamma}. 
To create independence between the new state and the others;
that is, between the new function and the others, 
we can implement the function in the control loop structure. 
In our previous studies \cite{Nakagawa12, Nakagawa13}, 
we developed a software evolution process based on control loops. 
The process constructs a system by combining control loops, 
which facilitates the localization of the impact of changes 
in the corresponding control loops. 
The process also enables us to reduce the effort required for 
finding and revising changes, which is a costly task 
in the software evolution process.

\begin{algorithm}[t]
 \label{alg:newFunction}
 \KwResult{The next action is executed}
 \KwData{{\it knowledge}: Knowledge component; $event_{a}$: Event}
State {\it current} = {\it knowledge.getCurrentState()}\;
Transition {\it t = knowledge.getTransition(current, $event_{a}$)}\;
{\it t.action()}\;
 \caption{Finding and Executing the next action.}
\end{algorithm}

When the event converter starts, 
it creates a hash table for the state machine models. 
All of the transitions available from the original and new
state machine models 
are registered
in the hash table in the knowledge component.
The hash table is used by calling the {\it getTransition} method, 
as indicated in Algorithm \ref{alg:newFunction}, 
in the plan and execute components. 
The converter determines the next action based on the current state and 
incoming event.
The Knowledge class provides the 
{\it putTransition(s0, s1, action, $event_{a}$)} method 
for registering transitions.
%
%
A transition is registered as a four-tuple: 
the current state ({\it s0}), incoming event (${\it event_{a}}$), 
next state ({\it s1}), and function to be called ({\it $action$}). 
%

\section{Experiment}\label{sect:experiment}

\begin{figure}[t]
  \centering
  \includegraphics[width=6cm]{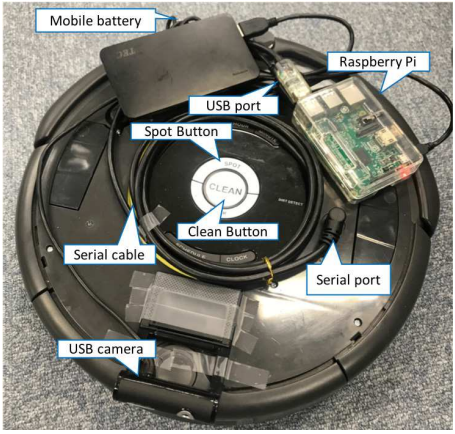}
  \caption{Cleaning robot using Raspberry Pi and USB camera.}
  \label{Figure:roomba}
\end{figure}

\begin{figure*}[t]
  \centering
  \includegraphics[width=13cm]{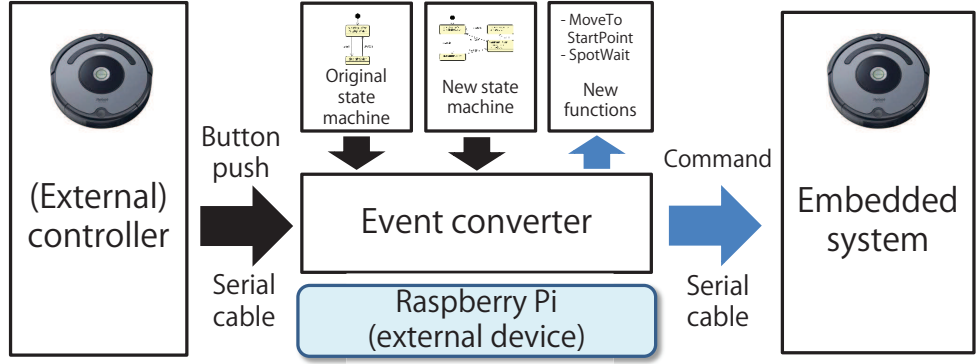}
  \caption{System architecture of cleaning robot used on the experiment.
The event converter extracted events sent from the cleaning robot. 
After determining the next action, 
the converter sent the events to the embedded system 
or invoked new functions.}
  \label{cleaning_overview}
\end{figure*}

\begin{figure}
  \centering
  \includegraphics[width=8cm]{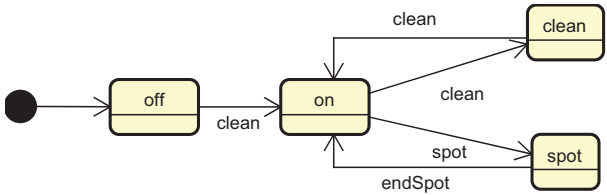}
  \caption{Original state machine model for cleaning robot.}
  \label{cleaning_original}
\end{figure}

To evaluate the proposed evolution mechanism, we conducted 
two experiments.
The first experiment handles an evolution of a cleaning robot in the real world (Exp. 1), and the second evaluates the performance and robustness
of the mechanism (Exp. 2).

\subsection{Exp. 1: Evolution of Cleaning Robot}
  
We applied the evolution mechanism to a cleaning robot 
as a real embedded system to evaluate its applicability and effectiveness.
According to the evolution process, we first defined 
the original and new state machine models for the cleaning robot. 
Thereafter, we implemented an event converter and new functions
using our framework and deployed them on an external device.
Finally, we verified whether the evolved robot behaved correctly 
by following the new state machine model.

In this experiment, we used a Roomba \cite{iRobot} 
as the cleaning robot and Raspberry Pi as the external device. 
The Roomba provides a serial port and serial interface 
\cite{RoombaInterface}. 
The event converter receives and sends events 
via a serial port provided by the Roomba. 
Figures \ref{Figure:roomba} and \ref{cleaning_overview} 
present a snapshot and the system architecture of the 
cleaning robot used in this experiment, respectively.
The cleaning robot has two main functions: {\it clean} and {\it spot}. 
When the CLEAN button provided by the robot is pressed, 
the robot navigates to clean the field automatically.
When the SPOT button is pressed, the robot intensely cleans 
a localized area by spiraling 
and then stops cleaning when it returns to the starting point. 
Figure \ref{cleaning_original} depicts the original state machine 
model for the cleaning robot. 
The robot cleans during the {\it clean} state. 
The {\it clean} state is changed to the {\it on} state 
when the robot receives a {\it clean} event invoked 
by pressing the button.
The robot spot cleans during the {\it spot} state. 
When the robot completes the spot cleaning, 
the state is changed to the {\it on} state by receiving the 
{\it endSpot} event sent by the robot itself.

We envision the following evolution scenario.
\begin{framed}
{\bf Evolution (move to the remote starting point):} 
The cleaning robot starts {\it spot} cleaning 
after it arrives at a starting point.
When the robot recognizes the starting point using a USB camera, 
the robot moves to that point.
\end{framed}

\subsubsection{Design}\label{subsect:experiment_design}
\begin{figure*}[htpb]
  \centering
  \includegraphics[width=16cm]{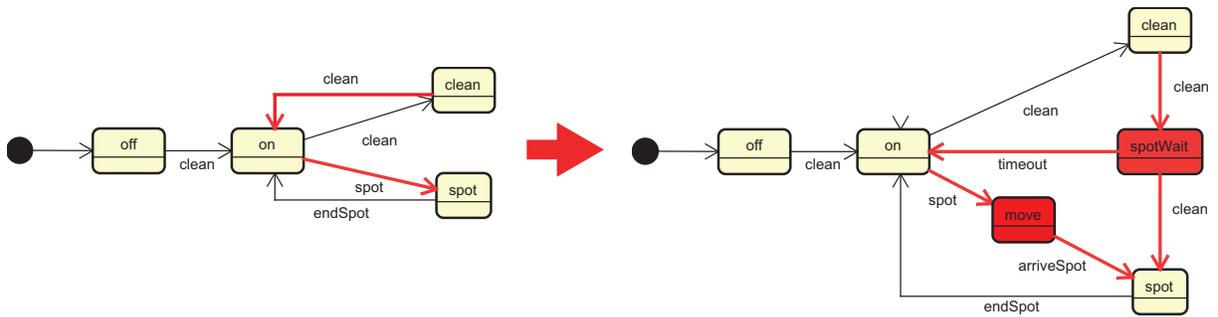}
  \caption{Changes in state machine model for cleaning robot.}
  \label{cleaning_original_to_new}
\end{figure*}

We begin with the definition of the original and new state machine models.
The new state machine model satisfies the new requirements described 
in the evolution scenario. 
In the design of the new system, we did not change existing states: 
the {\it off, on, spot,} and {\it clean} states. 
The system behavior of the robot was changed 
by adding new events and states. 
We used existing events provided by the cleaning robot, 
the {\it clean, spot} and {\it endSpot} events, 
without using additional buttons for new external events. 
To avoid making the existing functions unusable for mapping existing events to new states, we added new transitions to the existing states. 
This additional path enabled the preservation of the existing functions. 
We added two new states: {\it move} for moving to a starting point 
and {\it spotWait} for providing a detour path  
using the original {\it spot} mode. 
Figure \ref{cleaning_original_to_new} depicts the changes 
in the state machine model. 
The new event {\it timeout} enables the selection to start spot cleaning 
or to finish cleaning; 
the {\it arriveSpot} event informs the robot of the arrival.

\begin{eqnarray}
\label{eq:ourProcess4}
 {\it Sys_{n}} &=& ( {\it E_{n}, S_{n}, T_{n}} ) \nonumber \\
{\it  E_{n}} &=& \{ {\it clean, spot, endSpot, timeout, arriveSpot} \}  \nonumber \\
{\it S_{n}} &=& \{ {\it off, on, spot, clean, move, spotWait} \}  \nonumber \\
{\it T_{n}} &=& \{{\it ( off, clean )  \rightarrow on,} \nonumber \\
 && {\it ( on, spot) \rightarrow move,}  \nonumber \\ 
 && {\it ( on, clean) \rightarrow clean,}  \nonumber \\
 && {\it ( move, arriveSpot) \rightarrow spot,} \nonumber \\
 && {\it ( spot, endSpot ) \rightarrow on,} \nonumber \\
 && {\it ( clean, clean) \rightarrow spotWait,} \nonumber \\
 && {\it ( spotWait, timeout) \rightarrow on,} \nonumber \\
 && {\it ( spotWait, clean) \rightarrow spot} \}  \nonumber 
\end{eqnarray}

\begin{figure}[t]
  \begin{center}
  \includegraphics[width=8.4cm]{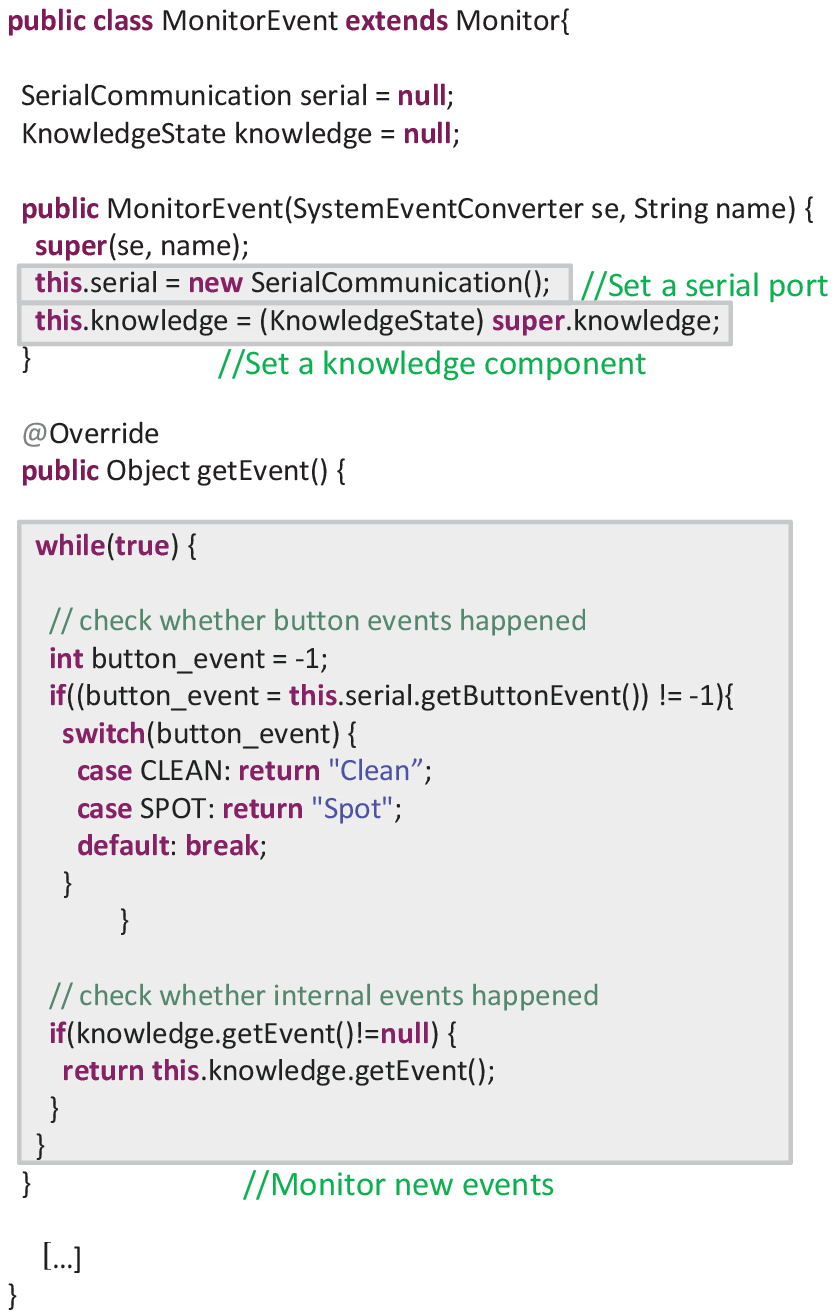}
  \caption{Part of MonitorEvent class.}
  \label{code:monitor}
  \end{center}
\end{figure}

\begin{figure}[t]
  \begin{center}
  \includegraphics[width=8.4cm]{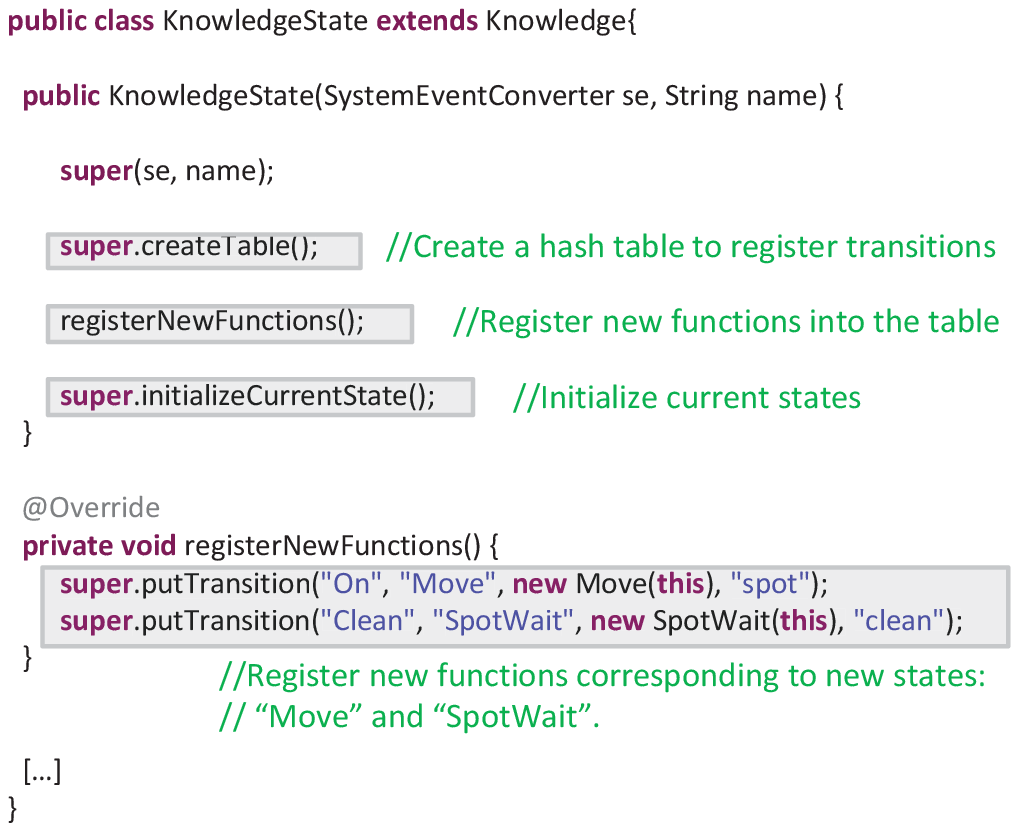}
  \caption{Part of KnowledgeState class.}
  \label{code:newfuctions}
  \end{center}
\end{figure}

%
%
%

\subsubsection{Implementation}
We developed an event converter using the proposed framework.
In particular, we implemented two classes for the converter: 
for monitoring events and sending events. 
The former is the {\it MonitorEvent} class that
inherits the {\it Monitor} super-class, 
as illustrated in Figure \ref{code:monitor}. 
The converter should detect the CLEAN and SPOT button events 
through the Roomba serial interface. 
Therefore, we used an external {\it SerialCommuniation} class 
for the Roomba serial interface. 
We implemented the event monitoring process
by overriding the {\it getEvent} method, 
which returns a new event when it arrives.

To implement new functions,
two new states, {\it move} and {\it spotWait} states, 
were added to the original state machine model in the design phase. 
We implemented two functions relating to the two respective states
and then registered the functions to the event converter 
such that the new functions can be called from the event converter.
We implemented the function relating to the {\it move} state 
using a control loop structure. 
The control loop mechanism provides high independence such that 
the functions can be separated from the other functions
and event converter. 
This mechanism is also beneficial for executing actions in parallel 
with less dependence. 
However, we implemented the function corresponding to the
{\it spotWait} state as a single class 
because the action is not complicated.
Figure \ref{code:newfuctions} presents the {\it KnowledgeState} class, 
which extends the {\it Knowledge} super-class. 
The constructor of the class has a transition hash table, 
the contents of which are automatically generated by parsing 
the state machine models. 
By overriding the {\it registerNewFunctions} method, 
new functions corresponding to the new states were registered to 
the knowledge component. 

\subsubsection{Results}
After deploying the event converter, 
new functions, and two state machine models on the external device,
we started the event converter using the commands listed 
in Table \ref{tbl:command}. 
Listings \ref{lst:1} and \ref{lst:2} present the logs 
of the cleaning robot execution. 
First, the CLEAN button was pushed (lines 1 to 13 in Listing \ref{lst:1}) 
to turn on the cleaning robot. 
When the SPOT button was pushed, the cleaning robot invoked the new function (lines 26 to 28) and moved to the marked point (lines 30 to 33), 
which was the expected behavior of the evolution scenario.
After the cleaning robot arrived at the starting point (lines 31 to 33), 
the robot suitably provided spot cleaning (lines 35 to 62). 
In the other execution (Listing \ref{lst:2}),
when the CLEAN button was pushed twice in the clean state (lines 1 and 18), 
the robot started the original spot cleaning (lines 28 to 33).
We observed that the new functions were successfully added 
to the cleaning robot without degrading functionality.

\lstset{ %
  frame=single,
  numbers=left,
  basicstyle={\scriptsize},
  backgroundcolor={\color{white}},   
  breaklines=true,                 
  captionpos=b,                    
  commentstyle=\color{mygreen},    
  escapeinside={\%*}{*)},          
  keywordstyle=\color{blue},       
  stringstyle=\color{mymauve},     
  xleftmargin=0.6cm,
  xrightmargin=0.5cm,
  stepnumber=1,
  showstringspaces=false,
  tabsize=1,
  breaklines=true,
  breakatwhitespace=false,
}
    

\begin{lstlisting}[
caption={Log of moving to remote starting point. This log \\
indicates that the robot executed a new function.}, 
label={lst:1}]
----   button_event : Clean   -----

Monitor
 inputs Clean event.
Analyze
 original_current_state: Off
 new_current_state: Off
 mode: Use existing functions
Plan
Execute
 MAPE-K loop will send this event : Clean
 original_current_state: On
 new_current_state: On

----   button_event : Spot   -----

Monitor
 inputs Spot event.
Analyze
 original_current_state: On
 new_current_state: On
 mode: Use new functions
Plan
 MAPE-K loop does not send events. 
Execute
 Operate existing functions for Move in the another thread
 original_current_state: On
 new_current_state: Move

***  Start to run new functions. ***
***  Arrive at target point  ***  
***  Send arriveSpot event to EventConverter.  ***  
***  Stop new functions  ***

----   internal_event : arriveSpot   -----

Monitor
 inputs arriveSpot event.
Analyze
 original_current_state: On
 new_current_state: Move
 mode: Use existing functions
Plan
Execute
 MAPE-K loop will send this event : Spot
 Spot command
 original_current_state: Spot
 new_current_state: Spot

----   internal_event : endSpot   -----

Monitor
 inputs endSpot event.
Analyze
 original_current_state: Spot
 new_current_state: Spot
 mode: Use existing functions
Plan
Execute
 MAPE-K loop will send this event : endSpot
 original_current_state: On
 new_current_state: On
\end{lstlisting}
\label{history1}

\begin{lstlisting}[
caption={Log of using spot cleaning mode. This log indicates that \\
the robot executed an original function following its evolution.}, 
label={lst:2}]
----   button_event : Clean   -----

Monitor
 inputs Clean event.
Analyze
 original_current_state: Clean
 new_current_state: Clean
 mode: Use new functions
Plan
 MAPE-K loop does not send events. 
Execute
 Operate existing functions for SpotWait in the another thread
 original_current_state: Clean
 new_current_state: SpotWait

***  Start to run new functions. ***

----   button_event : Clean   -----

Monitor
 inputs Clean event.
Analyze
 original_current_state: Clean
 new_current_state: SpotWait
 mode: Use new functions
Plan
Execute
 MAPE-K loop will send this event : Clean
 Clean command
 MAPE-K loop will send this event : Spot
 Spot command
 original_current_state: Spot
 new_current_state: Spot

\end{lstlisting}
\label{history2}


\subsection{Exp. 2: Performance and Robustness}

We evaluate the performance and robustness of the
proposed mechanism using the probabilistic model checking technique
\cite{Kwiatkowska:FSE07}.
Model checking is known as an effective technique for
developing critical applications \cite{FM:ACMServeys09}. 
Probabilistic model checking verifies models 
with state transitions annotated by probabilities or transition rates.
The technique determines whether QoS requirements specified in temporal logics are satisfied by the model.
In this study, we used PRISM \cite{PRISM4.0}, 
a probabilistic model checker,  for this experiment.
We constructed continuous-time Markov chain (CTMC) \cite{CTMC:Springer91} 
models and defined two QoS requirements:
\begin{itemize}
 \item {\it performance}: the process of the embedded system should 
finish as fast as possible.
 \item {\it robustness}: event loss caused by timing errors should be 
as small as possible. 
\end{itemize}

\begin{figure}[t]
  \begin{center}
  \includegraphics[width=8.4cm]{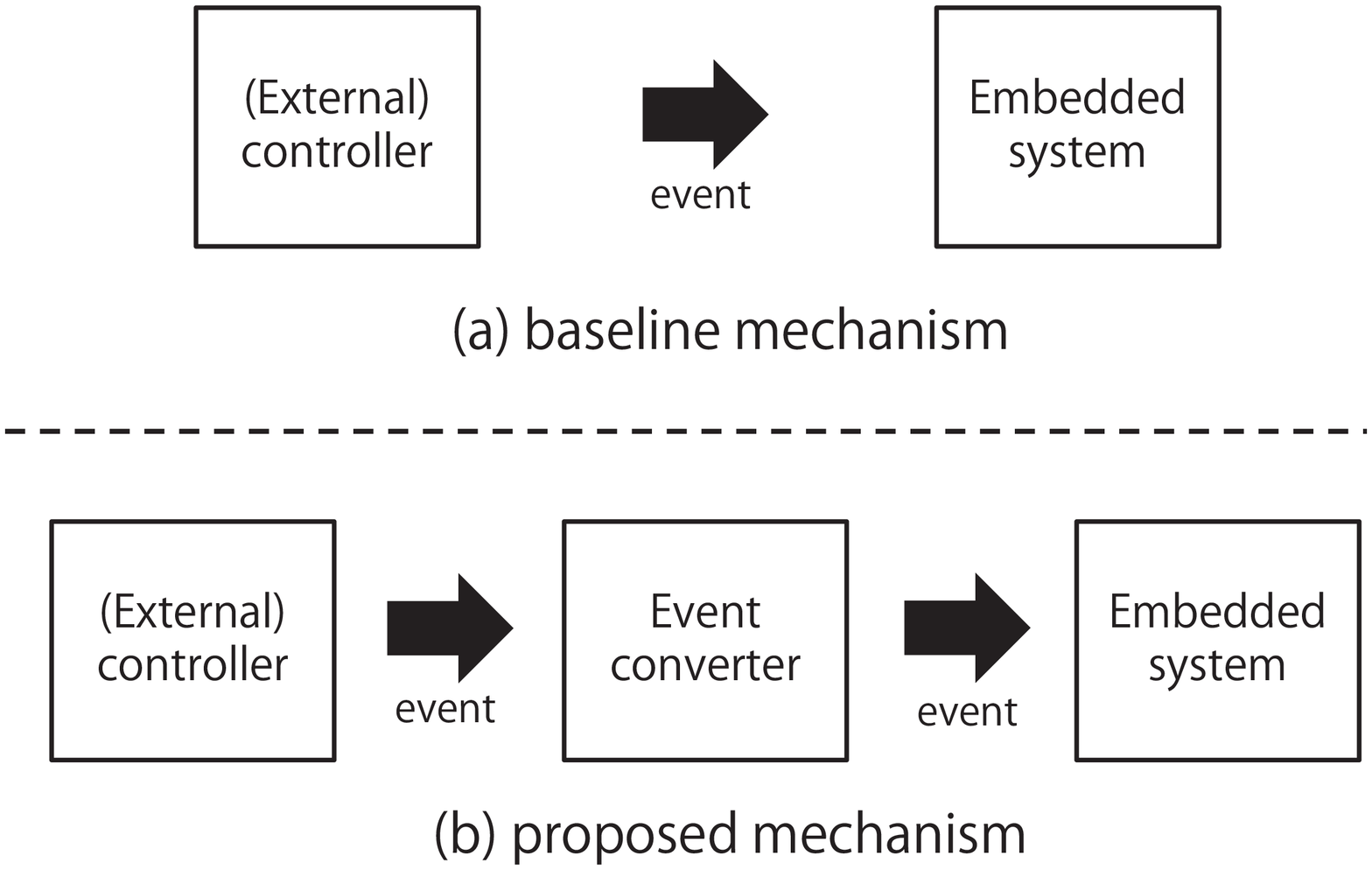}
  \caption{Two models used in Exp. 2. (a) is a baseline mechanism 
that corresponds to an ordinary embedded system.}
  \label{model:Exp2}
  \end{center}
\end{figure}

We adopted a modeling style used in various examples, 
such as \cite{952737}\cite{10.5555/3288647.3288709}.
Two CTMC models are used in Exp. 2; the first model is 
for the baseline mechanism, which corresponds to 
an ordinary embedded system, as illustrated in Figure \ref{model:Exp2} (a); 
the second model is
for the baseline mechanism, as illustrated in Figure \ref{model:Exp2} (b).
Listing \ref{code:prism-proposed} shows the CTMC model for 
the proposed mechanism, corresponding to model (b) 
in Figure \ref{model:Exp2}.

\begin{figure*}[t]
  \begin{center}
  \includegraphics[width=17.0cm]{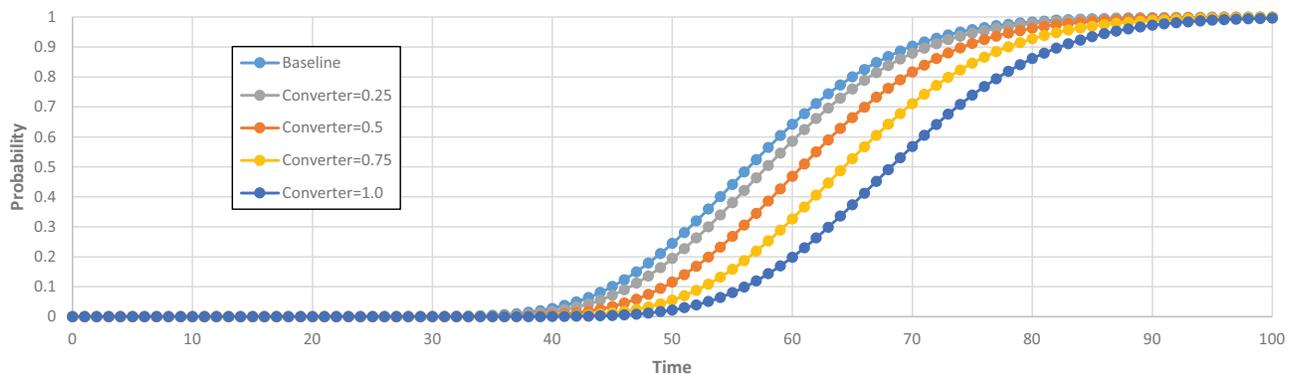}
  \caption{Performance evaluation results for Exp. 2.}
  \label{performance:Exp2}
  \end{center}
\end{figure*}

\begin{figure}[t]
  \begin{center}
  \includegraphics[width=8.4cm]{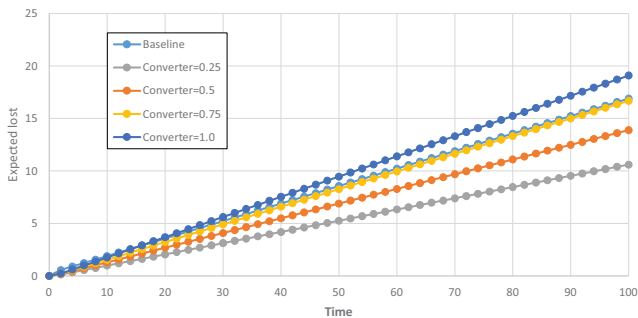}
  \caption{Event lost results for Exp. 2.}
  \label{lost:Exp2}
  \end{center}
\end{figure}

\begin{lstlisting}[
caption={The CTMC model for the proposed mechanism used in \\
Exp. 2. 
The controller illustrated in Figure \ref{model:Exp2} is not explicitly
described \\
in this model.}, 
label={code:prism-proposed}]
ctmc

//-------------------------------------------------
const int st_max = 20;

const double event_arrive = 1/2; // (mean inter-arrival time is 2 seconds)
const double emb_internal_process = 1/1; //(mean inter-arrival time is 1.0 seconds)
const double conv_internal_process = 1/0.25; //(mean inter-arrival time is 0.25 seconds)

module Converter
	arrived: bool init false;
	
	// conv_st = the number of arrived states of the event converter
	conv_st : [1..st_max] init 1;
	
	// A event arrives
	[arrived] arrived=false -> event_arrive : (arrived'=true); 
	[conv_lost] arrived=true -> event_arrive : (arrived'=true);

	//control embedded system
	[control] arrived -> conv_internal_process :  
		(conv_st'=min(conv_st+1,st_max))&(arrived'=false);

endmodule

module EmbeddedSystem
	emb_controlled : bool init false;
	lost : bool init false;
	
	// emb_st = the number of arrived states of the embedded system
	emb_st : [1..st_max] init 1;
	
	//event arrives
	[control] emb_controlled=false -> (emb_controlled'=true);
		
	[control] emb_controlled=true -> (lost'=true);
		
	[emb_lost] lost=true -> (lost'=false);

	//internal process has been finished
	[process] emb_controlled=true -> emb_internal_process : 
		(emb_st'=min(emb_st+1,st_max))& (emb_controlled'=false)& (lost'=false);
	
endmodule

//-------------------------------------------------

rewards "lost"
	[conv_lost] true: 1;
	[emb_lost] true: 1;
endrewards

\end{lstlisting}


The following parameters, all of which were defined 
in Listing \ref{code:prism-proposed}, were used in this experiment.
\begin{itemize}
 \item {\it st\_max} = 20 (line 4): State path length required to 
reach the final state.
 
 \item {\it event\_arrive} = 1/2 (line 6): Inverse of the average time interval between event arrival. It means that the average time interval between
event arrival is 2 s. 
 \item {\it emb\_internal\_process} = 1/1 (line 7): 
Inverse of the average action time of the embedded system. 

 \item {\it conv\_internal\_process} = 1/0.25, 1/0.5, 1/0.75, or 1/1 
(line 8): 
Inverse of the average conversion time. 
In this experiment, we used four values for the average conversion time: 
0.25, 0.5, 0.75, and 1 s. 
\end{itemize} 
Because
the average action time of the embedded system is 1 s, the average
conversion time represents
the ratio comparing with the action time of the embedded system.
For example, when the average conversion time is 0.5 s,
the converter is twice as fast as the embedded system.

To evaluate the two QoS requirements, we used the following two properties:

\begin{itemize}
 \item \verb|P=? [F [T,T] emb_st = st_max]|: This property calculates
the transient probability of the state of the embedded system being the final state (the 20th state). We can use this property for evaluating the performance of the mechanism. If the probability increases to 1.0 faster, 
the mechanism can reach the final state faster, i.e., the mechanism can finish tasks faster.

 \item \verb|R{"lost"}=? [C<=T]|: This property is used for the robustness evaluation. The \verb|R| operator represents the reward-based analysis.
In particular, \verb|C<=T| corresponds to the reward cumulated along a path until time \verb|T| has elapsed. 
Reward ``lost'' is defined in lines 48 to 51 
in Listing \ref{code:prism-proposed}. This reward increases when the converter or embedded system fails to receive events. 
Therefore, this property calculates the expected number of lost events.
We use this property for the robustness evaluation.
 
\end{itemize}

Figures \ref{performance:Exp2} and \ref{lost:Exp2} illustrate the results.
Figure \ref{performance:Exp2} shows the performance evaluation results
when changing the conversion time.
From this graph, we determined that the proposed mechanism provides similar
performances as the baseline mechanism when the conversion time is 
sufficiently faster than the action time of the embedded system
(converter=0.25 in the graph).
Figure \ref{lost:Exp2} shows the number of events lost.
From these results, 
we determined that the proposed mechanism improves the prevention of
the events lost when the conversion time is faster
than the action time of the embedded system.
%
These results show that the proposed mechanism
performs well and is sufficiently robust when the event conversion is
sufficiently faster than the action time of the embedded system.

\section{Discussion}\label{sect:discussion}

We discuss our evolution mechanism from 
design, applicability, implementation, and system evolution perspectives.

\subsection{Design}
Our evolution mechanism uses original and new state machine models 
to add new functions to an embedded system.
When the event converter determines that the next state exists
in the original model, the converter causes the embedded system 
to use the original function.
If the next state does not exist in the original model,
the converter executes a new function associated with 
the transition to the new state. 

The current framework can only handle flat state machine models,
which do not contain hierarchical or orthogonal structures.
We should translate a state machine model containing hierarchical 
or orthogonal structures into a flat model according to the equivalence
relationship described in \cite{statecharts}. 
The extension of the framework for considering the hierarchical and orthogonal structures is planned as the next step of our study.

The new state machine model must be carefully designed to avoid
introducing unexpected transitions and a lack of necessary 
transitions.
Introducing a mechanism to support 
the reachability analysis is beneficial at design time to not provoke 
system degradation. 
A model checking technique for verifying the correctness properties 
of state machine models can solve this problem.
This technique will aid in revising the models 
or disabling certain functions more safely.


Furthermore, the evolution mechanism uses an external device to deploy 
and execute new functions. 
When a new function uses an existing function implemented in
an embedded system, 
the state of the embedded system may change incorrectly.
This situation should be carefully considered.
In the cleaning robot scenario, the new function was able to
 be implemented without influencing the embedded system. 
However, when a new function changes the state of an embedded system, 
the new system must not change the state of the embedded system 
unintentionally. 
This problem can be solved in certain cases.
Two types of behaviors and their corresponding new states should be 
prepared: the states that affect the embedded system and 
states that recover the state of the embedded system. 
When implementing parts of new functions corresponding to 
the former states, 
we add additional features to maintain the current state of 
the embedded system. 
The latter states restore the state of the system 
after the new functions were executed and change the state of 
the embedded system.

Formal specifications can be applied in the design phase
to connect the design model with the implementation code.
For example, VDM++ \cite{Fitzgerald:2005:VDO:1044891} 
can describe specifications based on object-oriented design. 
VDM++ handles explicit descriptions that explain how functions are 
implemented and implicit descriptions that outline what is required in 
the functions. 
After designing the state machine models, these specifications may 
aid in conducting a stepwise refinement of the new functions.
Dom\'{i}nguez et al. \cite{10.1016/j.infsof.2012.04.008} provided
a systematic review of code generation from state machine specifications. 
The generation techniques described in the survey paper can be used; 
however, because we use state machine
models not to develop the system itself
but enable the event converter to identify the differences from 
the original behavior, the application of the technique will be partial.


\subsection{Applicability \& Implementation}
Although certain embedded systems have communication interfaces,
many embedded systems are closed from the outside. 
In this study, we assumed that embedded systems have communication 
interfaces, such as APIs or ports, to receive events 
from the event converter. 
In the cleaning robot scenario, a serial port was available for sending
events to the robot. 
However, when an embedded system does not provide any communication 
interfaces, a new mechanism must be constructed to send events
from the event converter. 
%
Even if an embedded system does not have communication interfaces, 
we can still control the system.
One solution is to develop a hardware device that affects the user
interfaces of the embedded system, such as a device that pushes
a button at the correct time.
Another solution is to change the environment of the embedded system.
For example, an air conditioner being operated when an external device
reduces or increases the temperature around the sensor of the air
conditioner.
These actions play the role of sending events to the embedded system.


Our framework does not restrict the implementation style of the new
functions in the evolution process. 
This enhances its affinity with other systems and applications, 
such as ROS2 \cite{ROS2}, which is the newest version of 
the Robot Operating System (ROS) \cite{ROS}.
Furthermore, this enables us to implement new features on 
additional devices.
For example, many external IoT devices can be used for this purpose.

The results of the cleaning robot experiment in Exp. 1 demonstrate 
the applicability of the evolution mechanism.
As shown in Exp. 2, the suitable event converter improves
the robustness of an embedded system by preventing event loss.
It is caused by the fast event converter successfully
receiving events from the controller. 
Because most embedded systems require a significant amount of time 
to provide their services, they tend to miss subsequent events. 
This is an advantage of the proposed mechanism.

Most embedded systems have a certain degree of 
time constraints.
First, the system performance must be considered.
%
The performance evaluation in Exp. 2 verifies that 
the proposed mechanism provides similar performance to 
a vanilla embedded system when the event conversion is sufficiently 
faster than the action of the embedded system.
Because the action time of an embedded system usually includes
the time to provide services using hardware, the event conversion time
is significantly faster than the time that the embedded system requires.
Second, strict time constraints imposed on an embedded system 
must be continuously observed.
Our current event converter does not contain mechanisms 
for verifying time constraints. 
We previously reported an initial study on a programming framework 
that handles time constraints at runtime \cite{NakagawaECRTS2019}.
The framework is for developing real-time systems and dynamically 
uses the model checking tool UPPAAL to verify time constraints. 
We can enhance the support for handling time constraints by
unifying this framework and the event converter.


\subsection{System Evolution}
When considering the system evolution, the system {\it maintainability} 
should be addressed.
In object-oriented development, a system is developed
by combining objects or components.
Object-oriented development generally improves maintainability; 
however, the amount of code tends to increase.
This characteristic opposes the requirement for embedded systems;
that is, software and hardware resources should be as small as possible.
%
Using an additional external device offers an advantage in this regard.
New functions can be implemented and executed without using limited 
hardware resources on the embedded system. 
%
The external approach also improves the maintainability 
of embedded systems. 
System evolution usually increases the complexity of 
the code; however, implementing new functions in the external
device maintains existing code from becoming more complex.
%
The comparison between the original and new state machine models helps
developers identify the differences visually. 
We proposed a software evolution technique \cite{Nakagawa13} 
that localizes changes into components using goal modeling.
This technique will also help separate new function modules 
from existing implemented functions.
The deployment support of our framework, that is, 
the commands for controlling new functions and their deployment, 
is also effective for system evolution.


\section{Related work}\label{sect:relatedwork}

This section summarizes related work in terms of design methods for 
embedded systems, reprogramming, firmware updating, frameworks using 
UML diagrams, and IoT system development.

\subsection{Design Methods for Embedded Systems}
\label{subsect:embeddedSystems}

%
%

Various design methods have been proposed for embedded systems.
Herrera et al. \cite{Herrera14} proposed the COMPLEX UML/MARTE design
space exploration methodology for embedded systems. 
This approach is based on a combination of model-driven engineering, 
the electronic system level, and design exploration technology. 
It uses the MAPERE profile, which offers a rich set of extensions 
specifically suited for the specification of 
embedded real-time systems. 
Lapalme et al. \cite{Lapalme:2006:NEE:1151074.1151082} proposed 
a .Net framework-based methodology for designing a new embedded system
design tool.
Apvrille and Roudier \cite{ApvrilleR14} defined a SysML-based 
model-driven development methodology for embedded systems,
which focuses on the security properties of embedded systems
in particular.
Heged\"{u}s et al. \cite{6100051} defined a model-driven framework
for design space exploration that analyzes implementation alternatives,
which satisfies all design constraints to identify the most suitable
design choice.
These studies handled the development of embedded systems;
however, the methods proposed in these studies were
assumed to be applied in the (initial) development of embedded systems,
and therefore, they
did not address the evolution of embedded systems.

A solution to the resource problem in embedded systems is 
an external approach that uses additional hardware.
%
Chung and Subramanian \cite{chung2001architecture} presented 
an architecture-based semantic evolution of embedded systems.
They focused on embedded systems that can be remotely controlled and 
identified four architecture types for this purpose: 
the 3Rs (rework, reload, and reboot), stored data, rule-based, 
and runtime module generation types. 
They used the NFR framework \cite{NFR} to select the suitable 
semantic evolution and its corresponding architecture type.
Berthier et al. \cite{Berthier13} proposed a global resource control 
approach based on a centralized view of the device states 
using a Boolean Mealy automata diagram.
Although their control layer was similar to our event converter,
they focused on the synchronous control of embedded systems to optimize 
global resource, and not the evolution of embedded systems.


\subsection{Reprogramming}\label{subsect:re-programming}
Reprogramming, which is the act of changing a program, 
is an effective method for adding new features to a system. 
In particular, the reprogramming of embedded systems is 
a long-standing problem in this field. 
Gay et al. \cite{pldi03gay} presented the nesC language, 
which supports the analysis of the system design 
by providing a programming model that incorporates event-driven 
execution and a flexible concurrency model. 
Furthermore, they developed the nesC compiler to perform 
whole-program analysis to reduce the resource consumption 
and improve the reliability. 
Shafi et al. \cite{6275799} proposed a reprogramming scheme 
consisting of a patch known as Queen's differential (QDiff). 
QDiff mitigates the effects of program layout changes and retains the maximum similarity between the old and new codes using similarity detection approaches. 
They focused on the lower-level problems of embedded system reprogramming,
such as power, speed, downtime, and reliability. 

Embedded systems are used extensively in IoT environments. 
Updating new features enables existing devices to be connected 
to one another when constructing IoT systems. 
The problem of adding new features to embedded systems 
is necessary for developing IoT systems 
and wireless sensor networks 
\cite{Dunkels:2006:RDL:1182807.1182810, Panta:2011:EIC:1921621.1921624, Sugihara:2008:PMS:1340771.1340774}. 
By reprogramming, attempts have been made to change the functionalities 
of the devices under resource constraints, 
such as energy, memory, and processing power, over time.

These approaches aid in reprogramming code implementations in embedded 
systems, whereas our evolution mechanism aims to ensure
that the existing systems are not modified. 
This mechanism enhances the modularity of existing systems. 
Both a reprogramming technique and programming framework can be used
in our evolution mechanism. 
The use of a reprogramming technique will help the framework 
communicate with the embedded system.


\subsection{Firmware}\label{subsect:firmware}
Firmware is software that controls hardware devices. 
Modern firmware is stored in EEPROM or flash memory, 
compared to old firmware architecture that was stored on ROM. 
Therefore, manufacturers can provide new features, 
revise bugs in their system, and protect users 
from security vulnerabilities through a firmware update. 
Such an update is performed via the USB or SD card and the Internet.

Mansor et al. \cite{Mansor15} analyzed the security of a firmware update 
protocol for vehicles.
Based on the results, they suggested several improvements 
in the protocols relating to safety and security measures. 
Jurkovic et al. \cite{Jurkovic14} proposed firmware update mechanisms 
for many distributed embedded devices controlled 
by a centralized server. 
Their approach provides dynamic upgrades of software in a rapid, 
robust, and reliable manner via the Internet.

Secure firmware updates are one of the most important issues 
in the IoT environment. 
Asokan et al. \cite{asokan2018assured} proposed a secure firmware update 
framework known as ASSURED, 
which provides a secure and scalable update for IoT systems. 
They considered realistic problems in large-scale IoT deployments 
while providing end-to-end security with enforceable constraints. 
Moreover, they decentralized heavy computational operations 
to external devices to place a minimal burden on the IoT devices.
Lee et al. \cite{Lee2017} proposed a blockchain-based secure firmware 
update methodology. 
When the firmware is updated, the embedded device downloads the latest 
software from a peer-to-peer firmware sharing blockchain network 
of nodes. 
Their approach allows vendors to provide new functionalities 
and patch vulnerabilities on embedded devices.

Taha and Mustafa \cite{Taha17} proposed a custom in-system firmware 
upgrading methodology using a serial peripheral interface (SPI).
They focused on large-scale embedded systems that do not have 
interfaces to run programming, such as the universal asynchronous 
receiver/transmitter. 
They updated the firmware through the SPI. 
In the cleaning robot example, we used the SPI provided by the robot 
when new features were added. 
However, we used it to send commands corresponding to an event
and not to upgrade the existing system.

The firmware ROS2 \cite{ROS2} provides libraries and tools 
to help software developers create robot applications 
such as hardware abstraction, device drivers, and package management. 
Furthermore, it provides many functionalities to create robots, 
owing to its open-software community and easy integration mechanisms 
based on self-programmed nodes. 
ROS enables robot applications to be created as well as the addition
of new features using libraries that strongly wrap the hardware system.

Firmware update techniques are used to add new features to 
embedded systems, whereas our evolution mechanism adds features 
without modifying existing systems. 
Neither technique conflicts with the other. 
Our mechanism can be used after a firmware update and new functions 
can be added using firmware. 
For example, our framework can remotely send events and commands via ROS. 
Furthermore, our framework, in which an implemented new function module 
uses a firmware such as ROS, can be used as a library.

\subsection{Development Using UML Diagrams}\label{subsect:midlware}

Numerous studies that have used UML diagrams have been conducted
in the software engineering field.
Yang \cite{Yang04, Yang05} outlined various methods for 
updating software systems using UML models and XML files. 
UML and XML are used as components of model-driven development
in various methods.
When the code of a software system is modified, 
refactoring \cite{Mens:TSE04} is generally conducted
to understand the code and localize changes. 
Similar to refactoring, {\it model refactoring} is used to restructure 
the design to improve its quality and reduce its complexity. 
Mens et al. \cite{Mens2007ModeldrivenSR} presented a model-driven software
development tool that uses UML diagrams, including class,
use case, and activity diagrams.
Sunye et al. \cite{Sunye:2001:RUM:647245.719454} presented 
a refactoring set and explained how the method can be designed 
to preserve the behavior of a UML model. 
Samek \cite{Samek08} presented a method for implementing 
the UML state machine model in C/C++. 
In general, UML diagrams cannot perform to their best potential 
without frameworks to implement the software systems 
efficiently based on these diagrams.
These frameworks enable systems to be designed and software to be
implemented more efficiently.

Kangas et al. \cite{Kangas:2006:UMS:1151074.1151077} proposed 
an automated UML-based design flow in which modeling and 
design control were handled in a single framework. 
They extended the UML 2.0 profile for back-annotating verification 
and architecture exploration, which was known as the TUT profile.
The information provided by the profile can aid a designer in making 
critical architectural decisions.
%
Riccobene et al. \cite{Riccobene:2009:SMD:1550987.1550993} extended 
UML 2.0 for SystemC and multithread C 
to provide a development methodology.
This approach helps software and hardware engineers improve 
the design flow of industrial embedded systems. 
The profile for SystemC enables the engineers to model resources 
and concurrency from a hardware perspective; 
the profile for multithread C allows engineers to model concurrency 
from a software perspective. 
Marinescu et al. \cite{Marinescu:2013:FSJ:2435227.2435241} presented 
FUSE for modeling and implementing embedded software components. 
This method provides a unified programming environment that supports 
statechart specifications and Java translations. 
By describing statecharts directly, FUSE prevents a lack of 
synchronization between the model and generated code. 
FUSE also enables tuning and debugging of the model and code 
within the same programming model. 

These techniques were proposed with frameworks for managing 
complex situations. 
The primary focus was on frameworks using diagrams that support 
the implementation phase of the initial development 
from the design phase. 
Our study complements this focus with the evolution aspect. 
Our method first revises an existing model and 
subsequently adds new functions to embedded systems using these models. 

\subsection{IoT System Development}
The IoT domain addresses issues focused on in this paper.
Several frameworks and middleware for IoT systems have been proposed. 
Armando et al. \cite{Armando:IoT21} proposed middleware that can 
handle heterogeneous IoT devices. 
Junejo et al. \cite{JunejoAisha:IoT21} 
proposed a framework for enhancing the security of the system, which provides a trust-based behavioral monitoring mechanism. 
Cheng et al. \cite{XieCheng:IoT21} introduced a knowledge graph-based multilayer IoT middleware. The middleware aims to solve the communication gap problem and heterogeneous access problem of IoT systems. The aim of these studies is similar to our study in terms of addressing device management and enhancement of the connectability of devices; however, our study mainly focuses on the function update of  devises, i.e., embedded systems.

Some studies have addressed the configuration and architecture of 
IoT systems.
Cai et al. \cite{MDDParternsforCoT:2018} proposed a rapid system 
development method using various service integration patterns. 
This method is based on Model-View-Controller (MVC) architecture.
Sun et al. \cite{SunG:IoTJ20} provided an energy-aware routing 
algorithm that minimizes 
server energy consumption while considering bandwidth consumption.
The algorithm is used for dynamic service function chain deployment.
Huang et al. \cite{HUANG2020107208} proposed 
a service-oriented network architecture 
to support the effective management of 5G-enabled IoT systems.
Bera et al. \cite{9222155} designed 
a blockchain-based IoT-enabled smart grid architecture.
We use an external approach for implementing embedded system evolution.
We currently assume that an event converter is used for
a target embedded system to be evolved; however,
when we handle several devices in an IoT system,
an event converter may be used for several devices.

Liu et al. \cite{LiuKaizheng:IoT21} proposed a manual reverse engineering 
framework for discovering the communication protocols of embedded Linux-based IoT systems. 
We assume that the original state machine model exists or can be described.
These original protocols and behavioral models are important for 
the appropriate incorporation of organizing IoT systems.
Cheng et al. \cite{ChengJiujun:IoT21} defined a dynamic evolution mechanism of Internet-of Vehicles (IoV) community. 
The mechanism uses a graph-based model to drive dynamic evolution. 
The evolution mechanism does not handle the implementation of components; 
however, components should have flexibility in terms of changes not 
preventing evolution. 
Our mechanism will aid such an evolution in enhancing the flexibility of components.

\section{Conclusions}\label{sect:conclustion}
To develop more fitting components for IoT systems,
this paper describes an evolution mechanism for 
updating the functionalities of embedded systems.
The mechanism uses a control unit, an event converter, 
which is deployed outside of the embedded system
to update the system without modifications.
A systematic evolution process and programming framework
helps implement the event converter.
Using the original and new state machine models, 
the event converter based on the MAPE-K loop structure appropriately
sends new events to the embedded system and executes new functions 
on the external device. 
%
The results of the first evaluation conducted on a cleaning robot 
demonstrated that our evolution mechanism 
can provide a model-based system design and 
API-based implementation to realize the evolution of an
embedded system without modification.
The second evaluation verified the performance and robustness
of the proposed evolution mechanism.
Note that the suitable event converter improves the robustness of
an embedded system by preventing event loss.

Future work includes the enhancement of the support for hierarchical and 
orthogonal state machine models.
Towards the model-driven engineering of secure and safe embedded systems, 
the introduction of a verification mechanism to ensure 
the accuracy of behaviors using model-checking techniques
is also planned.
We believe our mechanism contributes to 
the efficient and assured evolution
of embedded systems and development of IoT systems.



%

\section*{Acknowledgment}

This work was supported by the JSPS Grants-in-Aid for Scientific Research
(Grant Numbers 15K00097, 17KT0043, 20H04167), Telecommunications Advancement Foundation, and Asahi Glass Foundation. 

\ifCLASSOPTIONcaptionsoff
  \newpage
\fi



\bibliographystyle{IEEEtran}
\bibliography{myrefs}

\begin{thebibliography}{10}
\providecommand{\url}[1]{#1}
\csname url@samestyle\endcsname
\providecommand{\newblock}{\relax}
\providecommand{\bibinfo}[2]{#2}
\providecommand{\BIBentrySTDinterwordspacing}{\spaceskip=0pt\relax}
\providecommand{\BIBentryALTinterwordstretchfactor}{4}
\providecommand{\BIBentryALTinterwordspacing}{\spaceskip=\fontdimen2\font plus
\BIBentryALTinterwordstretchfactor\fontdimen3\font minus
  \fontdimen4\font\relax}
\providecommand{\BIBforeignlanguage}[2]{{%
\expandafter\ifx\csname l@#1\endcsname\relax
\typeout{** WARNING: IEEEtran.bst: No hyphenation pattern has been}%
\typeout{** loaded for the language `#1'. Using the pattern for}%
\typeout{** the default language instead.}%
\else
\language=\csname l@#1\endcsname
\fi
#2}}
\providecommand{\BIBdecl}{\relax}
\BIBdecl

\bibitem{6424332}
R.~Khan, S.~U. Khan, R.~Zaheer, and S.~Khan, ``Future internet: The internet of
  things architecture, possible applications and key challenges,'' in
  \emph{2012 10th International Conference on Frontiers of Information
  Technology}, Dec 2012, pp. 257--260.

\bibitem{Mens08}
T.~Mens and S.~Demeyer, \emph{Software evolution}.\hskip 1em plus 0.5em minus
  0.4em\relax Springer-Verlag Berlin Heidelberg, 2008.

\bibitem{Kephart03}
J.~O. Kephart and D.~M. Chess, ``The vision of autonomic computing,''
  \emph{IEEE Computer}, vol.~36, no.~1, pp. 41--50, 2003.

\bibitem{Oreizy99}
P.~Oreizy, M.~M. Gorlick, R.~N. Taylor, D.~Heimbigner, G.~Johnson,
  N.~Medvidovic, A.~Quilici, D.~S. Rosenblum, and A.~L. Wolf, ``An
  architecture-based approach to self-adaptive software,'' \emph{IEEE
  Intelligent Systems}, vol.~14, no.~3, pp. 54--62, 1999.

\bibitem{Shaw95}
M.~Shaw, ``Beyond objects: a software design paradigm based on process
  control,'' \emph{SIGSOFT Software Engineering Notes}, vol.~20, pp. 27--38,
  January 1995.

\bibitem{IBM05}
IBM, ``An architectural blueprint for autonomic computing third edition,''
  https://www-03.ibm.com/autonomic/pdfs{\slash}AC\%20Blueprint\%20White\%20Paper\%20V7.pdf.

\bibitem{Weyns13}
D.~Weyns, B.~Schmerl, V.~Grassi, S.~Malek, R.~Mirandola, C.~Prehofer,
  J.~Wuttke, J.~Andersson, H.~Giese, and K.~M. G{\"o}schka, ``On patterns for
  decentralized control in self-adaptive systems,'' in \emph{Software
  Engineering for Self-Adaptive Systems II}.\hskip 1em plus 0.5em minus
  0.4em\relax Springer, 2013, pp. 76--107.

\bibitem{porter2014smart}
M.~E. Porter and J.~E. Heppelmann, ``How smart, connected products are
  transforming competition,'' \emph{Harvard business review}, vol.~92, no.~11,
  p.~18, 2014.

\bibitem{TsuchidaCOMPSAC18}
S.~Tsuchida, H.~Nakagawa, E.~Tramontana, A.~Fornaia, and T.~Tsuchiya, ``A
  framework for updating functionalities based on the {MAPE} loop mechanism,''
  in \emph{2018 IEEE 42nd Annual Computer Software and Applications Conference
  (COMPSAC)}, vol.~01, July 2018, pp. 38--47.

\bibitem{UML}
O.~M. Group, ``{Unified Modeling Language (UML)},'' \url{https://www.uml.org/}.

\bibitem{Samek08}
M.~Samke, \emph{Practical UML Statecharts in C/C++: Event-Driven Programming
  for Embedded Systems}, 2008.

\bibitem{RumbaughBook91}
J.~Rumbaugh, M.~Blaha, W.~Premerlani, F.~Eddy, and W.~Lorensen,
  \emph{Object-Oriented Modeling and Design}.\hskip 1em plus 0.5em minus
  0.4em\relax Prentice-Hall, Inc., 1991.

\bibitem{Astah}
I.~Change~Vision, ``Astah,'' \url{http://astah.change-vision.com}.

\bibitem{raspi}
T.~R.~P. Foundation, ``Raspberry pi,'' https://www.raspberrypi.org/.

\bibitem{Tsuda15}
H.~Tsuda, H.~Nakagawa, and T.~Tsuchiya, ``Towards self-adaptation on real-world
  hardware: A preliminary lightweight programming framework,'' in \emph{2015
  IEEE 9th International Conference on Self-Adaptive and Self-Organizing
  Systems}, Sept 2015, pp. 176--177.

\bibitem{Nakagawa12}
H.~Nakagawa, A.~Ohsuga, and S.~Honiden, ``Towards dynamic evolution of
  self-adaptive systems based on dynamic updating of control loops,'' in
  \emph{2012 IEEE Sixth International Conference on Self-Adaptive and
  Self-Organizing Systems}, Sept 2012, pp. 59--68.

\bibitem{4221625}
J.~Kramer and J.~Magee, ``Self-managed systems: an architectural challenge,''
  in \emph{Future of Software Engineering (FOSE '07)}, May 2007, pp. 259--268.

\bibitem{Gamma}
E.~Gamma, \emph{Design patterns: elements of reusable object-oriented
  software}.\hskip 1em plus 0.5em minus 0.4em\relax Pearson Education India,
  1995.

\bibitem{Nakagawa13}
H.~Nakagawa, A.~Ohsuga, and S.~Honiden, ``A goal model elaboration for
  localizing changes in software evolution,'' in \emph{2013 21st IEEE
  International Requirements Engineering Conference (RE)}, July 2013, pp.
  155--164.

\bibitem{iRobot}
iRobot Corporation, ``{iRobot},'' https://global.irobot.com.

\bibitem{RoombaInterface}
iRobot corporation, ``Roomba interface,'' http://www.irobot.lv/\break
  uploaded\_files/File/iRobot\_Roomba\_500\_Open\_Interface\_Spec.pdf.

\bibitem{Kwiatkowska:FSE07}
\BIBentryALTinterwordspacing
M.~Kwiatkowska, ``Quantitative verification: Models techniques and tools,'' in
  \emph{Proc. of the the 6th Joint Meeting of the European Software Engineering
  Conference and the ACM SIGSOFT Symposium on The Foundations of Software
  Engineering (ESEC-FSE 2007)}, ser. ESEC-FSE '07.\hskip 1em plus 0.5em minus
  0.4em\relax New York, NY, USA: Association for Computing Machinery, 2007, p.
  449–458. [Online]. Available: \url{https://doi.org/10.1145/1287624.1287688}
\BIBentrySTDinterwordspacing

\bibitem{FM:ACMServeys09}
\BIBentryALTinterwordspacing
J.~Woodcock, P.~G. Larsen, J.~Bicarregui, and J.~Fitzgerald, ``Formal methods:
  Practice and experience,'' \emph{ACM Comput. Surv.}, vol.~41, no.~4, oct
  2009. [Online]. Available: \url{https://doi.org/10.1145/1592434.1592436}
\BIBentrySTDinterwordspacing

\bibitem{PRISM4.0}
M.~Kwiatkowska, G.~Norman, and D.~Parker, ``Prism 4.0: Verification of
  probabilistic real-time systems,'' in \emph{Computer Aided Verification},
  G.~Gopalakrishnan and S.~Qadeer, Eds.\hskip 1em plus 0.5em minus 0.4em\relax
  Berlin, Heidelberg: Springer Berlin Heidelberg, 2011, pp. 585--591.

\bibitem{CTMC:Springer91}
W.~J. Anderson, \emph{Continuous-time Markov chains : an applications-oriented
  approach}.\hskip 1em plus 0.5em minus 0.4em\relax Springer-Verlag, 1991.

\bibitem{952737}
Q.~Qiu, Q.~Qu, and M.~Pedram, ``Stochastic modeling of a power-managed
  system-construction and optimization,'' \emph{IEEE Transactions on
  Computer-Aided Design of Integrated Circuits and Systems}, vol.~20, no.~10,
  pp. 1200--1217, 2001.

\bibitem{10.5555/3288647.3288709}
S.~Gerasimou, R.~Calinescu, and G.~Tamburrelli, ``Synthesis of probabilistic
  models for quality-of-service software engineering,'' \emph{Automated
  Software Engineering.}, vol.~25, no.~4, p. 785–831, dec 2018.

\bibitem{statecharts}
D.~Harel, ``Statecharts: a visual formalism for complex systems,'' vol.~8, pp.
  231--274.

\bibitem{Fitzgerald:2005:VDO:1044891}
J.~Fitzgerald, P.~G. Larsen, P.~Mukherjee, N.~Plat, and M.~Verhoef,
  \emph{Validated Designs For Object-oriented Systems}.\hskip 1em plus 0.5em
  minus 0.4em\relax Santa Clara, CA, USA: Springer-Verlag TELOS, 2005.

\bibitem{10.1016/j.infsof.2012.04.008}
\BIBentryALTinterwordspacing
E.~Dom\'{i}nguez, B.~P\'{e}Rez, A.~L. Rubio, and M.~A. Zapata, ``A systematic
  review of code generation proposals from state machine specifications,''
  \emph{Information and Software Technology}, vol.~54, no.~10, p. 1045–1066,
  October 2012. [Online]. Available:
  \url{https://doi.org/10.1016/j.infsof.2012.04.008}
\BIBentrySTDinterwordspacing

\bibitem{ROS2}
O.~Robotics, ``Ros 2 documentation,'' https://docs.ros.org/en/foxy/index.html.

\bibitem{ROS}
M.~Quigley, K.~Conley, B.~P. Gerkey, J.~Faust, T.~Foote, J.~Leibs, R.~Wheeler,
  and A.~Y. Ng, ``Ros: an open-source robot operating system,'' in \emph{ICRA
  Workshop on Open Source Software}, 2009.

\bibitem{NakagawaECRTS2019}
H.~Nakagawa, H.~Tsuda, and T.~Tsuchiya, ``Towards real-time self-adaptation
  using a verification mechanism,'' in \emph{Proc. of the 31st Conference on
  Real-Time Systems (ECRTS’19), Work-in-progress}, 2019, pp. 10--12.

\bibitem{Herrera14}
\BIBentryALTinterwordspacing
F.~Herrera, H.~Posadas, P.~Pe\~{n}il, E.~Villar, F.~Ferrero, R.~Valencia, and
  G.~Palermo, ``The {COMPLEX} methodology for {UML/MARTE} modeling and design
  space exploration of embedded systems,'' \emph{Journal of Systems
  Architecture}, vol.~60, no.~1, pp. 55 -- 78, 2014. [Online]. Available:
  \url{http://www.sciencedirect.com/science/article/pii/S138376211300194X}
\BIBentrySTDinterwordspacing

\bibitem{Lapalme:2006:NEE:1151074.1151082}
\BIBentryALTinterwordspacing
J.~Lapalme, E.~M. Aboulhamid, and G.~Nicolescu, ``A new efficient {EDA} tool
  design methodology,'' \emph{ACM Transactions on Embedded Computing Systems},
  vol.~5, no.~2, pp. 408--430, May 2006. [Online]. Available:
  \url{http://doi.acm.org/10.1145/1151074.1151082}
\BIBentrySTDinterwordspacing

\bibitem{ApvrilleR14}
\BIBentryALTinterwordspacing
L.~Apvrille and Y.~Roudier, ``Towards the model-driven engineering of secure
  yet safe embedded systems,'' in \emph{Proceedings First International
  Workshop on Graphical Models for Security, GraMSec 2014, Grenoble, France,
  April 12, 2014.}, 2014, pp. 15--30. [Online]. Available:
  \url{https://doi.org/10.4204/EPTCS.148.2}
\BIBentrySTDinterwordspacing

\bibitem{6100051}
A.~Heged$\ddot{u}$s, A.~Horv\'{a}th, I.~R\'{a}th, and D.~Varr\'{o}, ``A
  model-driven framework for guided design space exploration,'' in \emph{2011
  26th IEEE/ACM International Conference on Automated Software Engineering (ASE
  2011)}, Nov 2011, pp. 173--182.

\bibitem{chung2001architecture}
L.~Chung and N.~Subramanian, ``Architecture-based semantic evolution: a study
  of remotely controlled embedded systems,'' in \emph{Proceedings IEEE
  International Conference on Software Maintenance. ICSM 2001}, Nov 2001, pp.
  663--666.

\bibitem{NFR}
L.~Chung, B.~A. Nixon, E.~Yu, and J.~Mylopoulos, \emph{Non-Functional
  Requirements in Software Engineering}.\hskip 1em plus 0.5em minus 0.4em\relax
  Springer, 2000.

\bibitem{Berthier13}
N.~Berthier, F.~Maraninchi, and L.~Mounier, ``Synchronous programming of device
  drivers for global resource control in embedded operating systems,''
  \emph{ACM Transactions on Embedded Computing Systems}, vol.~12, pp. 39:1--,
  03 2013.

\bibitem{pldi03gay}
D.~Gay, P.~Levis, R.~von Behren, M.~Welsh, E.~Brewer, and D.~Culler, ``{The
  nesC Language: A Holistic Approach to Network Embedded Systems},'' in
  \emph{{Proceedings of the ACM SIGPLAN 2003 Conference on Programming Language
  Design and Implementation (PLDI)}}, June 2003.

\bibitem{6275799}
N.~B. Shafi, K.~Ali, and H.~S. Hassanein, ``No-reboot and zero-flash
  over-the-air programming for wireless sensor networks,'' in \emph{2012 9th
  Annual IEEE Communications Society Conference on Sensor, Mesh and Ad Hoc
  Communications and Networks (SECON)}, June 2012, pp. 371--379.

\bibitem{Dunkels:2006:RDL:1182807.1182810}
\BIBentryALTinterwordspacing
A.~Dunkels, N.~Finne, J.~Eriksson, and T.~Voigt, ``Run-time dynamic linking for
  reprogramming wireless sensor networks,'' in \emph{Proceedings of the 4th
  International Conference on Embedded Networked Sensor Systems}, ser. SenSys
  '06.\hskip 1em plus 0.5em minus 0.4em\relax New York, NY, USA: ACM, 2006, pp.
  15--28. [Online]. Available: \url{http://doi.acm.org/10.1145/1182807.1182810}
\BIBentrySTDinterwordspacing

\bibitem{Panta:2011:EIC:1921621.1921624}
\BIBentryALTinterwordspacing
R.~K. Panta, S.~Bagchi, and S.~P. Midkiff, ``Efficient incremental code update
  for sensor networks,'' \emph{ACM Transactions on Sensor Networks}, vol.~7,
  no.~4, pp. 30:1--30:32, Feb. 2011. [Online]. Available:
  \url{http://doi.acm.org/10.1145/1921621.1921624}
\BIBentrySTDinterwordspacing

\bibitem{Sugihara:2008:PMS:1340771.1340774}
\BIBentryALTinterwordspacing
R.~Sugihara and R.~K. Gupta, ``Programming models for sensor networks: A
  survey,'' \emph{ACM Transactions on Sensor Networks}, vol.~4, no.~2, pp.
  8:1--8:29, Apr. 2008. [Online]. Available:
  \url{http://doi.acm.org/10.1145/1340771.1340774}
\BIBentrySTDinterwordspacing

\bibitem{Mansor15}
H.~Mansor, K.~Markantonakis, R.~N. Akram, and K.~Mayes, ``Don't brick your car:
  Firmware confidentiality and rollback for vehicles,'' in \emph{2015 10th
  International Conference on Availability, Reliability and Security}, Aug
  2015, pp. 139--148.

\bibitem{Jurkovic14}
G.~Jurkovic and V.~Sruk, ``Remote firmware update for constrained embedded
  systems,'' in \emph{2014 37th International Convention on Information and
  Communication Technology, Electronics and Microelectronics (MIPRO)}, May
  2014, pp. 1019--1023.

\bibitem{asokan2018assured}
\BIBentryALTinterwordspacing
N.~Asokan, T.~Nyman, N.~Rattanavipanon, A.~Sadeghi, and G.~Tsudik, ``{ASSURED:}
  architecture for secure software update of realistic embedded devices,''
  \emph{CoRR}, vol. abs/1807.05002, 2018. [Online]. Available:
  \url{http://arxiv.org/abs/1807.05002}
\BIBentrySTDinterwordspacing

\bibitem{Lee2017}
\BIBentryALTinterwordspacing
B.~Lee and J.-H. Lee, ``Blockchain-based secure firmware update for embedded
  devices in an internet of things environment,'' \emph{The Journal of
  Supercomputing}, vol.~73, no.~3, pp. 1152--1167, Mar 2017. [Online].
  Available: \url{https://doi.org/10.1007/s11227-016-1870-0}
\BIBentrySTDinterwordspacing

\bibitem{Taha17}
M.~A.~A. Taha and S.~Mustafa, ``Custom in-system firmware upgrade for {MSP430
  Microcontrollers family using SPI},'' in \emph{2017 International Conference
  on Communication, Control, Computing and Electronics Engineering (ICCCCEE)},
  Jan 2017, pp. 1--5.

\bibitem{Yang04}
H.~Yang, \emph{Software Evolution with UML and XML}.\hskip 1em plus 0.5em minus
  0.4em\relax IGI Global, 2004.

\bibitem{Yang05}
------, \emph{Advances in UML and XML-based software evolution}.\hskip 1em plus
  0.5em minus 0.4em\relax IGI Global, 2005.

\bibitem{Mens:TSE04}
T.~Mens and T.~Tourw\'{e}, ``A survey of software refactoring,'' \emph{IEEE
  Transactions on Software Engineering}, vol.~30, no.~02, pp. 126--139, feb
  2004.

\bibitem{Mens2007ModeldrivenSR}
T.~Mens, G.~Taentzer, and D.~M{\"u}ller, ``Model-driven software refactoring,''
  in \emph{Proc. of the Workshop on Refactoring Tools (WRT 2007)}, 2007.

\bibitem{Sunye:2001:RUM:647245.719454}
\BIBentryALTinterwordspacing
G.~Suny{\'e}, D.~Pollet, Y.~L. Traon, and J.-M. J{\'e}z{\'e}quel, ``Refactoring
  uml models,'' in \emph{Proceedings of the 4th International Conference on The
  Unified Modeling Language, Modeling Languages, Concepts, and Tools}, ser.
  \&\#171;UML\&\#187; '01.\hskip 1em plus 0.5em minus 0.4em\relax Berlin,
  Heidelberg: Springer-Verlag, 2001, pp. 134--148. [Online]. Available:
  \url{http://dl.acm.org/citation.cfm?id=647245.719454}
\BIBentrySTDinterwordspacing

\bibitem{Kangas:2006:UMS:1151074.1151077}
\BIBentryALTinterwordspacing
T.~Kangas, P.~Kukkala, H.~Orsila, E.~Salminen, M.~H\"{a}nnik\"{a}inen, T.~D.
  H\"{a}m\"{a}l\"{a}inen, J.~Riihim\"{a}ki, and K.~Kuusilinna, ``{UML-based}
  multiprocessor {SoC} design framework,'' \emph{ACM Transactions on Embedded
  Computing Systems}, vol.~5, no.~2, pp. 281--320, May 2006. [Online].
  Available: \url{http://doi.acm.org/10.1145/1151074.1151077}
\BIBentrySTDinterwordspacing

\bibitem{Riccobene:2009:SMD:1550987.1550993}
\BIBentryALTinterwordspacing
E.~Riccobene, P.~Scandurra, S.~Bocchio, A.~Rosti, L.~Lavazza, and
  L.~Mantellini, ``Systemc/c-based model-driven design for embedded systems,''
  \emph{ACM Transactions on Embedded Computing Systems}, vol.~8, no.~4, pp.
  30:1--30:37, Jul. 2009. [Online]. Available:
  \url{http://doi.acm.org/10.1145/1550987.1550993}
\BIBentrySTDinterwordspacing

\bibitem{Marinescu:2013:FSJ:2435227.2435241}
\BIBentryALTinterwordspacing
M.-C. Marinescu and C.~S\'{a}nchez, ``Fusing statecharts and java,'' \emph{ACM
  Transactions on Embedded Computing Systems}, vol.~12, no.~1s, pp.
  45:1--45:21, Mar. 2013. [Online]. Available:
  \url{http://doi.acm.org/10.1145/2435227.2435241}
\BIBentrySTDinterwordspacing

\bibitem{Armando:IoT21}
N.~Armando, J.~Sá~Silva, and F.~Boavida, ``An approach to the unified
  management of heterogeneous iot environments,'' \emph{IEEE Internet of Things
  Journal}, vol.~8, no.~8, pp. 6916--6927, 2021.

\bibitem{JunejoAisha:IoT21}
A.~K. Junejo, N.~Komninos, and J.~A. McCann, ``A secure integrated framework
  for fog-assisted internet-of-things systems,'' \emph{IEEE Internet of Things
  Journal}, vol.~8, no.~8, pp. 6840--6852, 2021.

\bibitem{XieCheng:IoT21}
C.~Xie, B.~Yu, Z.~Zeng, Y.~Yang, and Q.~Liu, ``Multilayer internet-of-things
  middleware based on knowledge graph,'' \emph{IEEE Internet of Things
  Journal}, vol.~8, no.~4, pp. 2635--2648, 2021.

\bibitem{MDDParternsforCoT:2018}
H.~Cai, Y.~Gu, A.~V. Vasilakos, B.~Xu, and J.~Zhou, ``Model-driven development
  patterns for mobile services in cloud of things,'' \emph{IEEE Transactions on
  Cloud Computing}, vol.~6, no.~3, pp. 771--784, 2018.

\bibitem{SunG:IoTJ20}
G.~Sun, R.~Zhou, J.~Sun, H.~Yu, and A.~V. Vasilakos, ``Energy-efficient
  provisioning for service function chains to support delay-sensitive
  applications in network function virtualization,'' \emph{IEEE Internet of
  Things Journal}, vol.~7, no.~7, pp. 6116--6131, 2020.

\bibitem{HUANG2020107208}
\BIBentryALTinterwordspacing
M.~Huang, A.~Liu, N.~N. Xiong, T.~Wang, and A.~V. Vasilakos, ``An effective
  service-oriented networking management architecture for 5g-enabled internet
  of things,'' \emph{Computer Networks}, vol. 173, p. 107208, 2020. [Online].
  Available:
  \url{https://www.sciencedirect.com/science/article/pii/S1389128619311442}
\BIBentrySTDinterwordspacing

\bibitem{9222155}
B.~Bera, S.~Saha, A.~K. Das, and A.~V. Vasilakos, ``Designing blockchain-based
  access control protocol in iot-enabled smart-grid system,'' \emph{IEEE
  Internet of Things Journal}, vol.~8, no.~7, pp. 5744--5761, 2021.

\bibitem{LiuKaizheng:IoT21}
K.~Liu, M.~Yang, Z.~Ling, H.~Yan, Y.~Zhang, X.~Fu, and W.~Zhao, ``On manually
  reverse engineering communication protocols of linux-based iot systems,''
  \emph{IEEE Internet of Things Journal}, vol.~8, no.~8, pp. 6815--6827, 2021.

\bibitem{ChengJiujun:IoT21}
J.~Cheng, C.~Cao, M.~Zhou, C.~Liu, S.~Gao, and C.~Jiang, ``A dynamic evolution
  mechanism for iov community in an urban scene,'' \emph{IEEE Internet of
  Things Journal}, vol.~8, no.~9, pp. 7521--7530, 2021.

\end{thebibliography}

\end{document}